\documentclass[aps,twocolumn,superscriptaddress,showpacs,10pt,longbibliography]{revtex4-1}
\usepackage{graphicx}
\usepackage{hyperref}
\usepackage{amssymb}
\usepackage{color}

\newcommand{\mup}{$\mu^+$}
\newcommand{\Mup}{Mu$^+$}
\newcommand{\Mu}{Mu$^0$}
\newcommand{\Mun}{Mu$^-$}
\newcommand{\MuT}{Mu$_{\rm T}$}
\newcommand{\MuTn}{Mu$_{\rm T}^0$}
\newcommand{\MuTm}{Mu$_{\rm T}^-$}
\newcommand{\MuBC}{Mu$_{\rm BC}$}
\newcommand{\MuBCn}{Mu$_{\rm BC}^0$}
\newcommand{\MuBCp}{Mu$_{\rm BC}^+$}

\begin{document}


\title{
Direct observation of hole carrier density profiles and their light induced manipulation at the surface of Ge
}



\author{T.~Prokscha}
\email{thomas.prokscha@psi.ch}
\affiliation{Laboratory for Muon Spin Spectroscopy, Paul Scherrer Institut, CH-5232 Villigen PSI, Switzerland}
\author{K-H.~Chow}
\affiliation{Department of Physics, University of Alberta, Edmonton T6G 2E1, Canada}
\author{Z.~Salman}
\affiliation{Laboratory for Muon Spin Spectroscopy, Paul Scherrer Institut, CH-5232 Villigen PSI, Switzerland}
\author{E.~Stilp}
\affiliation{Laboratory for Muon Spin Spectroscopy, Paul Scherrer Institut, CH-5232 Villigen PSI, Switzerland}
\affiliation{Physics Institute, University of Zurich, 8057 Zurich, Switzerland}
\author{A.~Suter}
\affiliation{Laboratory for Muon Spin Spectroscopy, Paul Scherrer Institut, CH-5232 Villigen PSI, Switzerland}

\date{\today}

\begin{abstract}
We demonstrate that, by using low-energy positive muon ($\mu^+$) spin spectroscopy as a local probe technique,
the profiles of free charge carriers can be directly determined in the accumulation/depletion surface regions of
p- or n-type Ge wafers. The detection of free holes is accomplished by measuring the effect of the
interaction of the free carriers with the $\mu^+$ probe spin on the observable muon spin polarization. 
By tuning the energy of the low-energy $\mu^+$ between 1 keV and 20 keV the near-surface region between 
10 nm and 160 nm is probed. 
We find hole carrier depletion and electron accumulation in all samples with doping concentrations
up to the $10^{17}$ cm$^{-3}$ range, which is opposite to the properties of cleaved Ge surfaces. 
By illumination with light the hole carrier density in the depletion zone
can be manipulated in a controlled way. Depending on the used light wavelength $\lambda$ this change can be 
persistent ($\lambda = 405, 457$~nm) or non-persistent ($\lambda = 635$~nm) at temperatures $< 270$~K. 
This difference is attributed to the different kinetic energies of the photo-electrons. Photo-electrons
generated by red light do not have sufficient energy to overcome
a potential barrier at the surface to be trapped in empty surface acceptor states.
Compared to standard macroscopic transport measurements 
our contact-less local probe technique
offers the possibility of measuring carrier depth profiles and manipulation directly.
Our approach may provide important microscopic information on a nanometer scale in semiconductor device studies.
%
\end{abstract}
%
%
%
%
%
%
\maketitle
%
%
%
%
%
\section{Introduction}\label{introduction}
The characterization of semiconductor materials and devices is key for understanding and developing semiconductor technologies. Specifically, the changes and the controlled manipulation
of charge carrier concentrations at semiconductor interfaces are of fundamental importance for their functionality in devices. With the tremendous growth of the field of
semiconductor device physics and the advancement of experimental characterization techniques over the past decades an enormous progress has been achieved, providing new insights and
improvements of semiconductor devices \cite{sze_physics_2006,schroder_semiconductor_2006,ferry_semiconductors_2013}.

Usually, a combination of macroscopic transport measurements and simple modeling are used to determine charge carrier depth profiles and electric field gradients across semiconductor 
interfaces \cite{sze_physics_2006,schroder_semiconductor_2006}. Carrier densities can be determined by Hall effect measurements, without depth resolution.
The density of dopants can be determined by optical techniques (infrared spectroscopy, photoluminescence, plasma resonance, and free carrier absorption) where depth resolution 
is also limited. Capacitance-voltage measurements are usually used
to gain information about free carrier and impurity profiles in a non-destructive way. These measurements require 
the manufacture of a Schottky contact, which is not always easy to build. 
The depth resolution of this technique is limited by the zero-bias
space-charge region at the surface, by voltage breakdown at larger depths, and by the 
Debye limit \cite{schroder_semiconductor_2006}.
%

A local probe technique, capable of detecting the variation of carrier densities as a function of depth, offers the unique possibility of measuring carrier profiles and
their manipulation directly. Here, we employ a beam of polarized low-energy positive muons (LE-\mup) with tuneable energies between 1 and 20~keV, and implant the \mup~at 
variable mean depths
between 10~nm and 120~nm in commercial Ge wafers. In semiconductors and insulators the \mup~stops at an intersitial site, where it can capture zero, one or two electrons
to form the hydrogen-like muonium states \Mup, \Mu, or \Mun~\cite{patterson_muonium_1988,cox_muonium_2009}. The interaction of free charge carriers with these muonium
states may cause a detectable change of muon spin polarization, which can be observed by measuring the time evolution of the \mup~polarization in a muon spin rotation 
($\mu$SR) experiment \cite{patterson_muonium_1988,cox_muonium_2009,yaouanc_muon_2011}. In this way, the stopped \mup~may act as ``sensor'' for free charge carriers.
The technique of using LE-\mup~in depth selective low-energy $\mu$SR (LE-$\mu$SR)
has been recently applied to measure the persistent photo-induced inversion of a Ge surface layer from n- to p-type \cite{prokscha_photo-induced_2013}, 
and the effect of band bending on the activation energy of shallow muonium states close to the surface of commercial CdS and ZnO wafers \cite{prokscha_depth_2014}.
These experiments demonstrate the capability of LE-$\mu$SR to determine quantitatively i), the charge carrier concentrations in a near-interface region in certain cases
(which was in Ref.~\cite{prokscha_photo-induced_2013} a photo-induced hole carrier concentration of $\sim 1.5\times 10^{14}$~cm$^{-3}$)
and ii), the electric field profile due to band bending at a semiconductor surface/interface \cite{prokscha_depth_2014}.

In this paper we significantly advance the methodology to perform depths scans of carrier concentrations of n- and p-type Ge wafers. We directly explore by means
of LE-$\mu$SR the hole depletion region width at the surface of p-type Ge, and we demonstrate
the manipulation of the hole carrier concentration $p$ 
in the depletion region by using illumination with a blue LED light source or laser (wavelengths $\lambda = 405, 457$~nm), or a red laser ($\lambda = 635$~nm). 
Whereas after illumination with blue light a persistent increase of hole carrier density
$p \gg 10^{14}$~cm$^{-3}$ is observed in the depletion zone, i.e. filling of the depletion region, illumination with red light induces a dynamic charge carrier equilibrium
with $p \sim 1\times10^{12}$~cm$^{-3}$ at a depth of 20~nm, increasing to $\sim 4\times10^{12}$~cm$^{-3}$ at a depth of 120~nm in a p-type Ge wafer with nominal 
$p \sim 10^{15}$~cm$^{-3}$. At the same time, electron accumulation is found in the surface region without illumination at depths of order 100~nm.
Our method has the potential to provide new insights in charge carrier transport phenomena on a nanometer scale
at semiconductor interfaces, relevant for device technology.
%
%
%
%
%
%
\section{Effect of free charge carriers on muon spin polarization}\label{EffectFreeCarriers}
Implantation of a positively charged muon (\mup) in a semiconductor or insulator normally leads to the formation of a hydrogen-like
muonium (Mu) state, where the \mup~may capture one or two electrons to form the neutral \Mu~or negatively charged \Mun~state, or it
ends up without capturing an electron (\Mup). Thus,
analogous to hydrogen, the three charge states \Mup, \Mu, and \Mun~can occur, depending on their formation energies and free charge
carrier concentrations. The two states \Mup~and \Mun, where the \mup~is not coupled to an unpaired electron, are called \textit{diamagnetic}
states, in contrast to the \textit{paramagnetic} state \Mu, where the hyperfine coupling with the bound unpaired electron causes an
additional magnetic field on the \mup. In a transverse-field muon spin rotation experiment (TF-\mup SR), where an external magnetic field $B$
is applied perpendicular to the initial muon spin direction, the muon spin in the \textit{diamagnetic} states precesses at the muon Larmor 
frequency $\nu_\mu = \gamma_\mu/(2\pi)\cdot B$ [$\gamma_\mu/(2\pi) = 135.54$~MHz/T, where $\gamma_\mu$ is the muon gyromagnetic ratio],
whereas in \Mu, much higher frequencies corresponding to transitions between various hyperfine states are observed \cite{yaouanc_muon_2011}.
 
In the group IV or III-V semiconductors with cubic diamond or zinc blende crystal structure, two Mu sites are known: Mu at a
bond center (\MuBC), or Mu at the tetrahedral interstitial site (\MuT) \cite{patterson_muonium_1988,cox_muonium_2009}. The \MuBC~state is usually the
donor-like configuration with positive or neutral charge (\MuBCp, \MuBCn), whereas the acceptor like \MuT~configuration can be in the
neutral or negative charge state (\MuTn, \MuTm). The electron distribution of the \MuBCn~state is axially anisotropic, with symmetry axis along
the $\langle 111\rangle$ direction. The electron density is centred at the host atoms, and relatively low at the \mup~site. This explains the small
value of the hyperfine coupling $A_{hfc}$ of \MuBCn~in Ge of the order of 100~MHz, compared to vacuum muonium with $A_{hfc} = 4.46$~GHz. In contrast, 
\begin{figure*}[ht]
\includegraphics[width=0.49\linewidth]{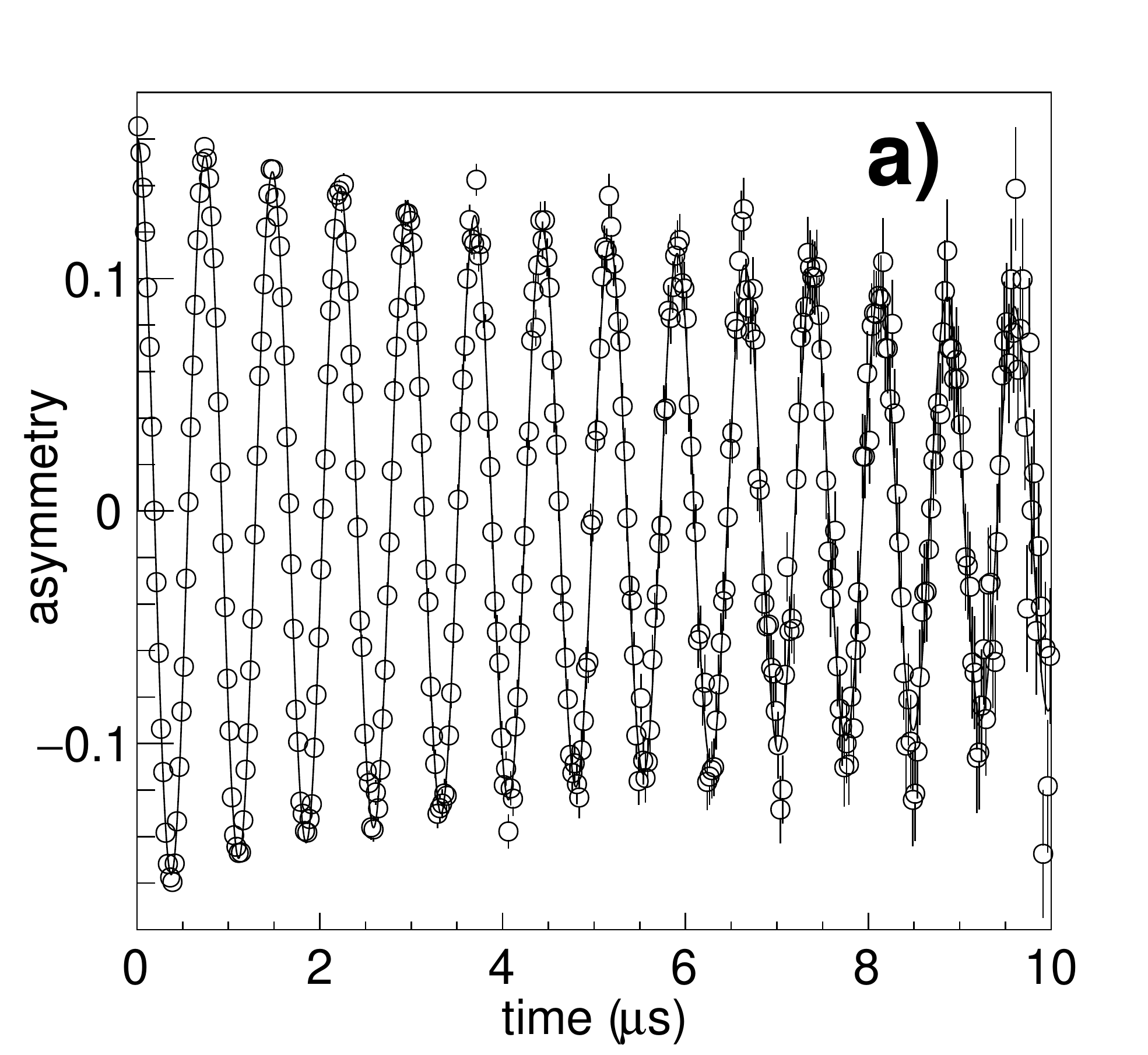}%
\includegraphics[width=0.49\linewidth]{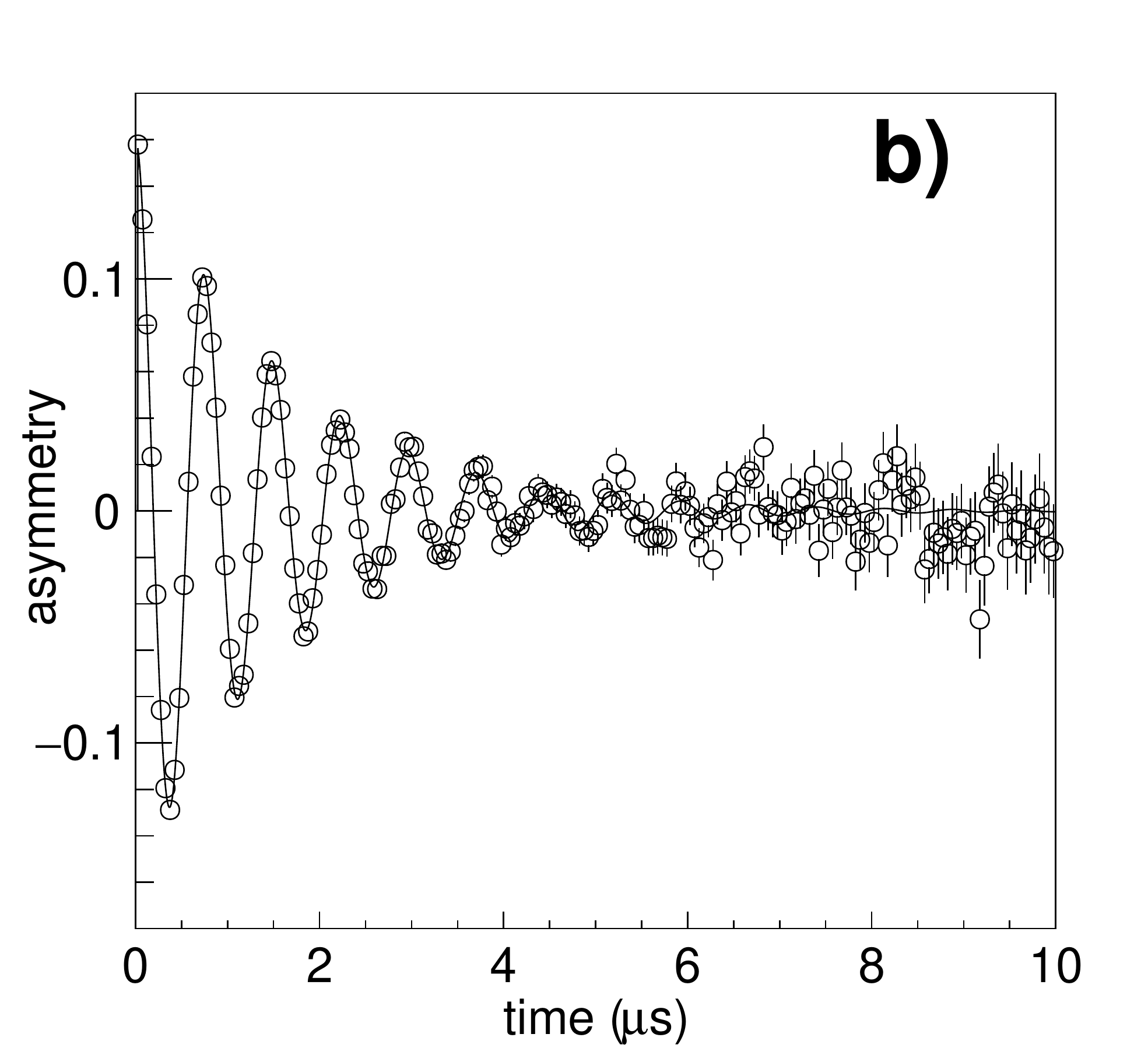}
\includegraphics[width=0.49\linewidth]{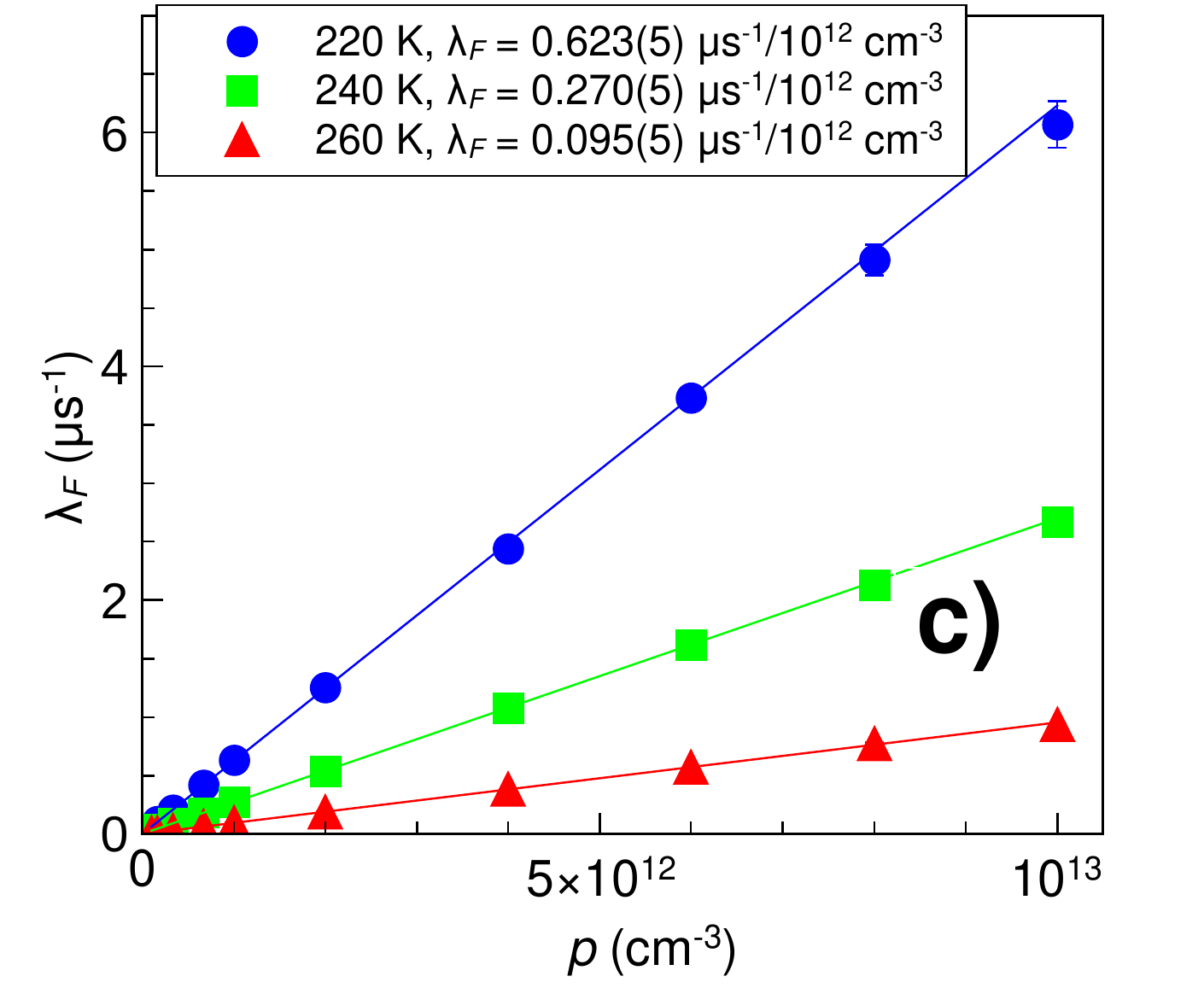}%
\includegraphics[width=0.49\linewidth]{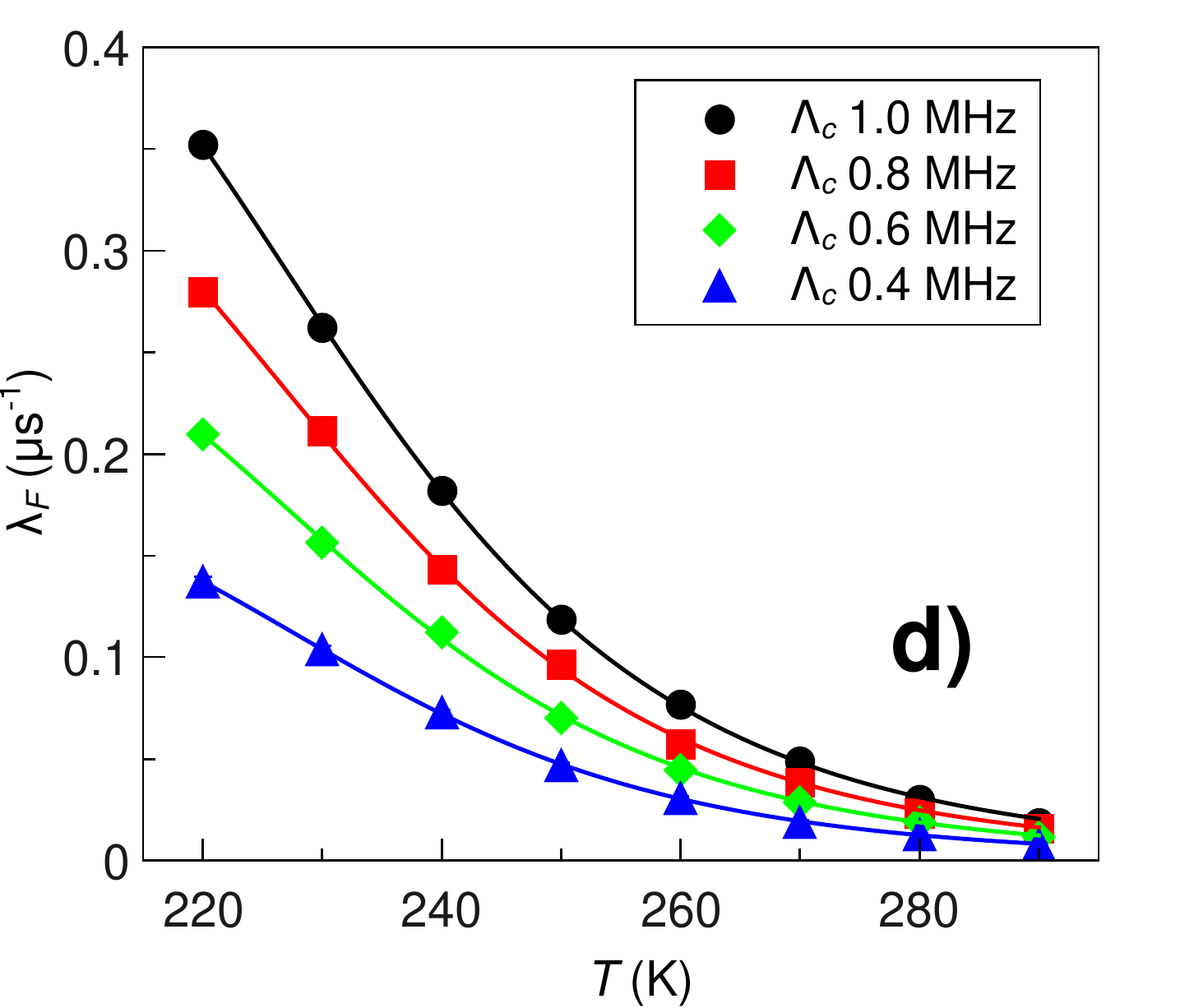}
\caption{Simulation data in a tranverse field of 10~mT of the charge-cycle of Eq.~\ref{eq1}, where we used for the \MuTn~ionization $E_A = 154$~meV and $\Lambda_0 = 3.2\times 10^{13}$~s$^{-1}$ (appendix~\ref{appendixI}). a) and b) are for $T = 220$~K, with hole capture rates $\Lambda_c = 0.18$~MHz and 1.8~MHz, corresponding to $p = 10^{11}$~cm$^{-3}$ and $p = 10^{12}$~cm$^{-3}$ (appendix~\ref{appendixI}), respectively. c) Fast depolarization rate $\lambda_{\rm F}$ as a function of hole carrier concentration $p$ and at various temperatures. The solid lines are linear fits to the data, with the slopes given in the legend. d) $\lambda_{\rm F}$ as a function of temperature for different hole capture rates $\Lambda_c$. The solid lines are fits of Eq.~\ref{eq2}.}\label{Fig1}
\end{figure*}
the electron density of \MuTn~in Ge is much larger and isotropic, with $A_{hfc} = 2.36$~GHz at low temperatures ($T < 40$~K), and linearly decreasing with
temperature at $T > 40$~K due to the coupling of \MuT~with phonons \cite{holzschuh_direct_1983, lichti_charge-state_1999}.

The sensitity of muons to free charge carriers originates from the interaction of the charged/neutral muonium states with these carriers. The interaction may lead 
to a change of the diamagnetic and paramagnetic fractions, or to a change of muon spin depolarization rates in the presence of cyclic charge-exchange processes, or to
phase shifts in the muon spin precession signal if a neutral pre-cursor state transforms into a diamagnetic state \cite{patterson_muonium_1988,cox_muonium_2009,  yaouanc_muon_2011}. The low temperature charge fractions in undoped Ge are
about 75\% in \MuTn, about 10\%-20\% in \MuBCn, and less than 10\% in a diamagnetic state \cite{patterson_muonium_1988,fan_influence_2008}. In the presence of
free electrons with concentration $n > 10^{17}$~cm$^{-3}$ in doped samples the 
\MuTn~acceptor state may capture an electron to form the diamagnetic \MuTm~state,
resulting in an increase of the diamagnetic fraction at low temperatures to 30\% at $n \sim 2\times 10^{18}$~cm$^{-3}$, and to 80\% at 
$n \sim 2\times 10^{19}$~cm$^{-3}$ \cite{andrianov_interaction_1978}. Increasing the temperature in Ge leads first to an onset of thermally activated ionization of
the \MuBCn~state at 150~K with an activation energy of about 145~meV \cite{lichti_hydrogen_2008}, where the \MuBCn~is completely transformed
to \MuBCp~at 200~K \cite{fan_influence_2008}. At $T > 200$~K thermally activated ionization of \MuTn$ \rightarrow$ \MuTm~sets in with an activation energy
of about 170~meV \cite{cox_muonium_2009,fan_influence_2008,prokscha_simulation_2014}, where the diamagnetic fraction reaches 100\% close to room temperature. 

It is the \MuTn~state at $T > 200$~K that we use as a sensor for free hole carriers. In the presence of free holes a recombination of a hole
$h^+$ with a \MuTm~state may occur to form again the neutral state. As a consequence charge-exchange cycles between \MuTn~and \MuTm~are established. These
lead to a depolarization of the TF-$\mu$SR precession signal due to fast turning on/off of the hyperfine field in the \MuTn~state, with a depolarization rate
proportional to the hole carrier concentration $p$ \cite{fan_influence_2008,prokscha_photo-induced_2013,prokscha_simulation_2014}. In the absence of holes the depolarization
rate of the diamagnetic signal is slow, with rate $\lambda_S < 0.2$~$\mu$s$^{-1}$ at $T < 300$~K \cite{prokscha_photo-induced_2013}. In the presence of charge-cycles,
a fast component with depolarization rate $\lambda_F > \lambda_S$ appears, where $\lambda_F \propto \Lambda_c$ with $\Lambda_c$ the hole capture rate.

We developed a Monte-Carlo simulation code \cite{prokscha_monte-carlo_2012,prokscha_simulation_2014} for the charge-exchange cycles which allows us, in combination with a calibration measurement on a p-doped wafer with $p = 10^{15}$~cm$^{-3}$, the determination of the hole capture rate $\Lambda_c$ for a measured muon spin depolarization rate $\lambda_F$ \cite{prokscha_simulation_2014}. The simulation is modeling the cyclic reaction
\begin{equation}
 \rm{Mu}_{T}^{0} \rightleftharpoons \rm{Mu}_{T}^{-} + h^{+}, \label{eq1}
\end{equation}
where the forward reaction (\MuTn~ionization) is described by an Arrhenius rate process with the temperature dependent ionization rate $\Lambda_i(T) = \Lambda_0\exp(-E_A/k_B T)$, $\Lambda_0$ the attempt frequency, $E_A$ the activation energy, and $k_B$ the Boltzmann constant. The reverse reaction, hole capture of \MuTm, is governed by the hole capture rate $\Lambda_c(T) = p\cdot (v_h \sigma_c^h)(T)$, where we assume a constant 
$p$ in the temperature range of the experiment, while the temperature dependence of $\Lambda_c(T)$ is absorbed in the product of the hole carrier velocity $v_h$ and the hole capture cross section $\sigma_c^h$. It has been shown in the calibration experiment that $v_h \sigma_c^h \propto T^{-2.2(2)}$ (see appendix~\ref{appendixI}), indicating that the temperature dependence of $v_h \sigma_c^h$ is goverend by the temperature dependence of the hole mobility ($\propto T^{-2.3}$) \cite{prokscha_simulation_2014}.

In Ge the assignment of the diamagnetic state at $T > 200$~K has been a matter of debate. It was proposed that the diamagnetic state is \MuBCp, which should then also be 
involved in cyclic charge-exchange reactions \cite{lichti_charge-state_1999,kadono_evidence_1997}. However, 
the \MuBCp~state cannot explain the observed suppression of diamagnetic fraction in the presence of holes (Fig.~\ref{Fig3} and \cite{fan_influence_2008}). The observed temperature dependence of $v_h \sigma_c^h \propto T^{-2.2(2)}$ points towards the involvement of holes in the charge cycles \cite{prokscha_simulation_2014}, which also favors the \MuTm~as diamagnetic state. Additionally,
if \MuBCp~were the relevant diamagnetic state, a large electron concentration of order $10^{17}$~cm$^{-3}$ should result in a suppression of diamagnetic fraction, or in the appearance of a fast component in this temperature range. However, the investigated sample with $6\times 10^{17}$~cm$^{-3}$ electron doping in Fig.~\ref{Fig3} a) does neither show a suppression of diamagnetic
fraction, nor a fast component. We take all these observations as strong evidence, 
that the relevant diamagnetic state is \MuTm~at $T > 200$~K, and that it is the
charge cycle of Eq.~\ref{eq1} determining the muon spin dynamics.

Figure~\ref{Fig1} displays the simulation results in an applied transverse field of 10~mT. The increase of the depolarization rate with increasing hole capture rate is obvious from the $\mu$SR asymmetry spectra in Fig.~\ref{Fig1}(a) and (b). Figure~\ref{Fig1}(c) demonstrates the linear relationship between hole carrier density - which determines the hole capture rate - and the emergent fast depolarization rate $\lambda_{F}$ in the presence of charge cycles. Figures~\ref{Fig1}(c) and (d) show, that for a given $p$, $\lambda_{F}$ decreases with increasing temperature. This is due to the exponentially increasing \MuTn~ionization rate, which means that the muons spend less and less time in the neutral state, causing less and less dephasing/depolarization of the muon spin during the charge cycles. The solid lines in Fig.~\ref{Fig1}(d) are fits of the equation 
\begin{equation}
  \lambda_{F} = 0.5\cdot \frac{\Lambda_i\Lambda_c}{\Lambda_i+\Lambda_c} \cdot \frac{\omega_0^2}{\Lambda_i^2+\omega_0^2(1+x^2)}, \label{eq2}
\end{equation}
where $\omega_0 = 2\pi\cdot 2150$~MHz is the hyperfine coupling of \MuTn~averaged over the temperature
range 220~K -- 290~K \cite{lichti_charge-state_1999}, and $x = B/B_0$ with $B_0 = \omega_0/(\gamma_\mu-\gamma_e) = 0.0764$~T is the hyperfine magnetic field at the electron, with gyromagnetic ratio $\gamma_e$. Equation~\ref{eq2} is the expression for 1/$T_1$, where 1/$T_1$ is the depolarization rate 
in a longitudinal field (field applied parallel to
the initial muon spin direction, LF-$\mu$SR) \cite{lichti_charge-state_1999,chow_diffusion_1996,chow_muonium_1993}. Due to the fact that in Ge the TF-$\mu$SR depolarization rate caused by nuclear dipolar fields is small compared to $\lambda_{\rm F}$, the TF-$\mu$SR depolarization rate $\lambda_{\rm F}$ can be well approximated by 1/$T_1$. This is further supported by the fact that the TF-$\mu$SR simulated data in Fig.~\ref{Fig1} d) are very well fitted
by Eq.~\ref{eq2}.
Equation~\ref{eq2} will be used in the analysis of the depth dependent hole carrier profiles under illumination. The temperature dependence of $\lambda_F$ is mainly determined by the exponential temperature dependence of $\Lambda_i$, and to a lesser extent by $\Lambda_c$ and $\omega_0$. With increasing temperature, $\Lambda_i$ is exponentially increasing, causing the decrease of $\lambda_F$.  
\section{Experimental Details}\label{ExperimentalDetails}
The samples were commercial Ge (100) 2'' wafers, 0.5-mm-thick, and nominally undoped, n- or p-type with doping ranges between $4\times 10^{14}$~cm$^{-3}$ and 
$6\times 10^{17}$~cm$^{-3}$, with relative uncertainties of 10\% for n-type and 20\% for p-type wafers, as specified by the suppliers
(MTI Corporation, Richmond CA, United States, and Crystec GmbH, Berlin, Germany).
The undoped sample was n-type with a doping of $\sim 5\times 10^{13}$~cm$^{-3}$,
as specified by the supplier.

The LE-$\mu$SR experiments were carried out at the low-energy muons (LEM) facility at the $\mu$E4 beam line\cite{prokscha_new_2008} of the Swiss Muon Source
(S$\mu$S, Paul Scherrer Institut, Villigen, Switzerland).
\begin{figure}[ht]
\includegraphics[width=1.0\linewidth]{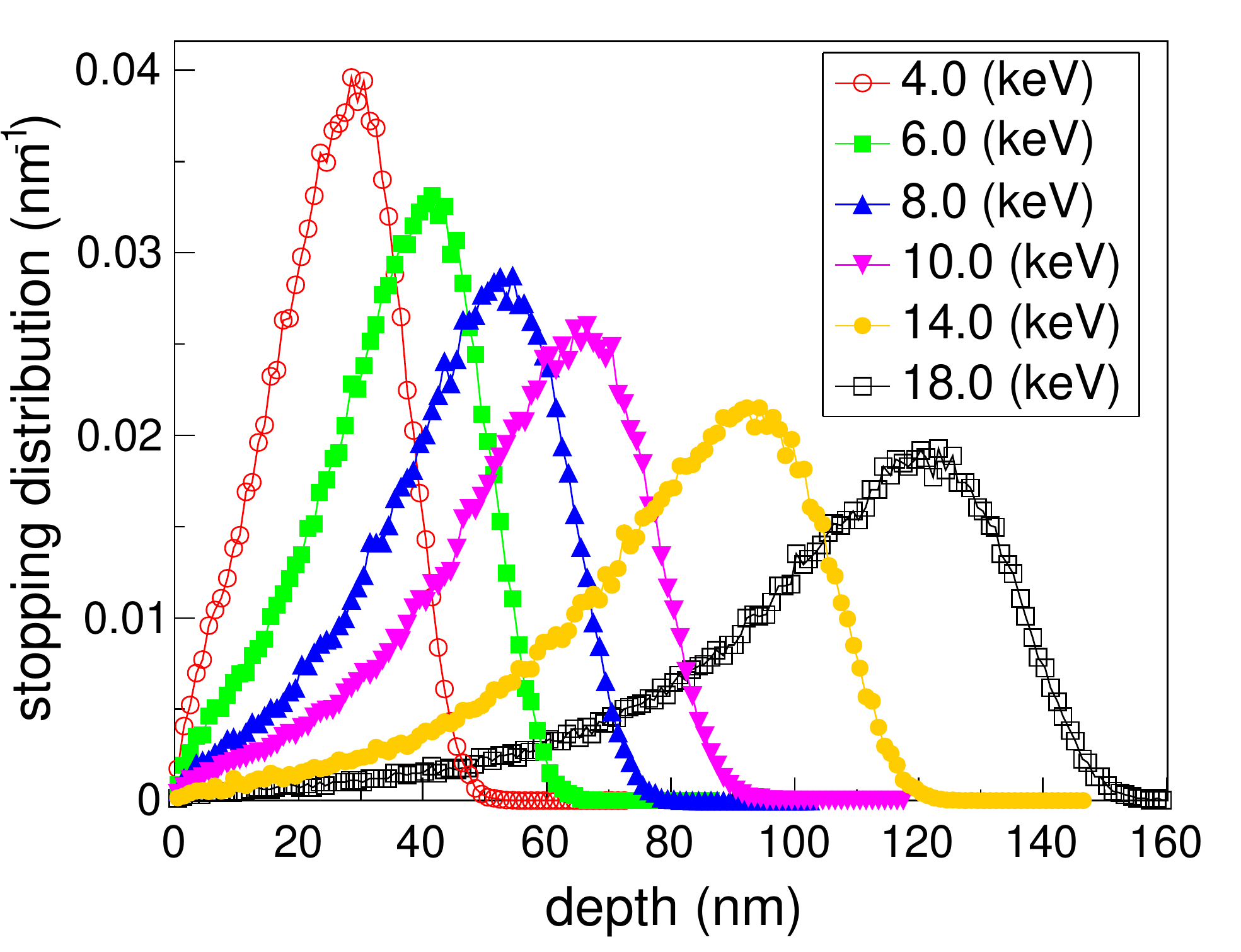}
\caption{Calculated muon stopping profiles for various implantation energies in Ge using the 
 program \tt{TrimSP}\cite{trimsp,morenzoni_implantation_2002}.}\label{Fig2}
\end{figure}
A beam of polarized $\mu^+$ with keV energies is generated by moderating a 4-MeV-$\mu^+$ beam, generated by the PSI proton accelerator, in a cryogenic moderator layer of solid Ar/N$_2$\cite{morenzoni_generation_1994,morenzoni_low-energy_2000,prokscha_moderator_2001}. The moderated muons with eV energies are electrostatically accelerated up to 20~keV and transported by electrostatic elements to the sample region. The samples are glued with conductive silver paint on a Ag-coated sample plate made of aluminum, where the final implantation energy $E_{imp}$ is adjusted by applying an electric potential up to $\pm 12.5$~kV to the sample plate. The implantation profiles of muons in Ge with energies between 4 and 18~keV are displayed in Fig.~\ref{Fig2}. To illuminate the samples, either LEDs or solid state lasers are available \cite{prokscha_low-energy_2012,prokscha_photo-induced_2013,stilp_controlling_2014}. For illumination with blue light we used either a LED source with $\lambda = 405$~nm (Bluepoint, H\"onle AG, Gr\"afeling, Germany) or a diode pumped solid state laser with $\lambda = 457$~nm. For red light a diode laser with $\lambda = 635$~nm was used (both lasers from DelMar Photonics Inc San Diego CA, United States). In all cases the maximum light intensity at the sample was 50 - 100 mW/cm$^2$.    

In addition to the near-surface investigations with LE-$\mu^+$, the bulk of the Ge wafers was studied at a mean depth of about 300~$\mu$m employing the two instruments DOLLY and GPS \cite{amato_new_2017} which use 4-MeV-$\mu^+$. The original wafers were cut in about 1~cm$^2$ pieces to fit into the cryostats of the 
bulk-$\mu$SR instruments. In all muon experiments the magnetic field was applied parallel to the $\langle 100 \rangle$ direction, and transverse to the initial muon spin direction.

%
%
%
\section{Results}\label{SecResults}
\subsection{Diamagnetic fractions at different depths and doping levels}\label{SubSecDiamagnetic}
\begin{figure*}[ht]
\includegraphics[width=0.495\linewidth]{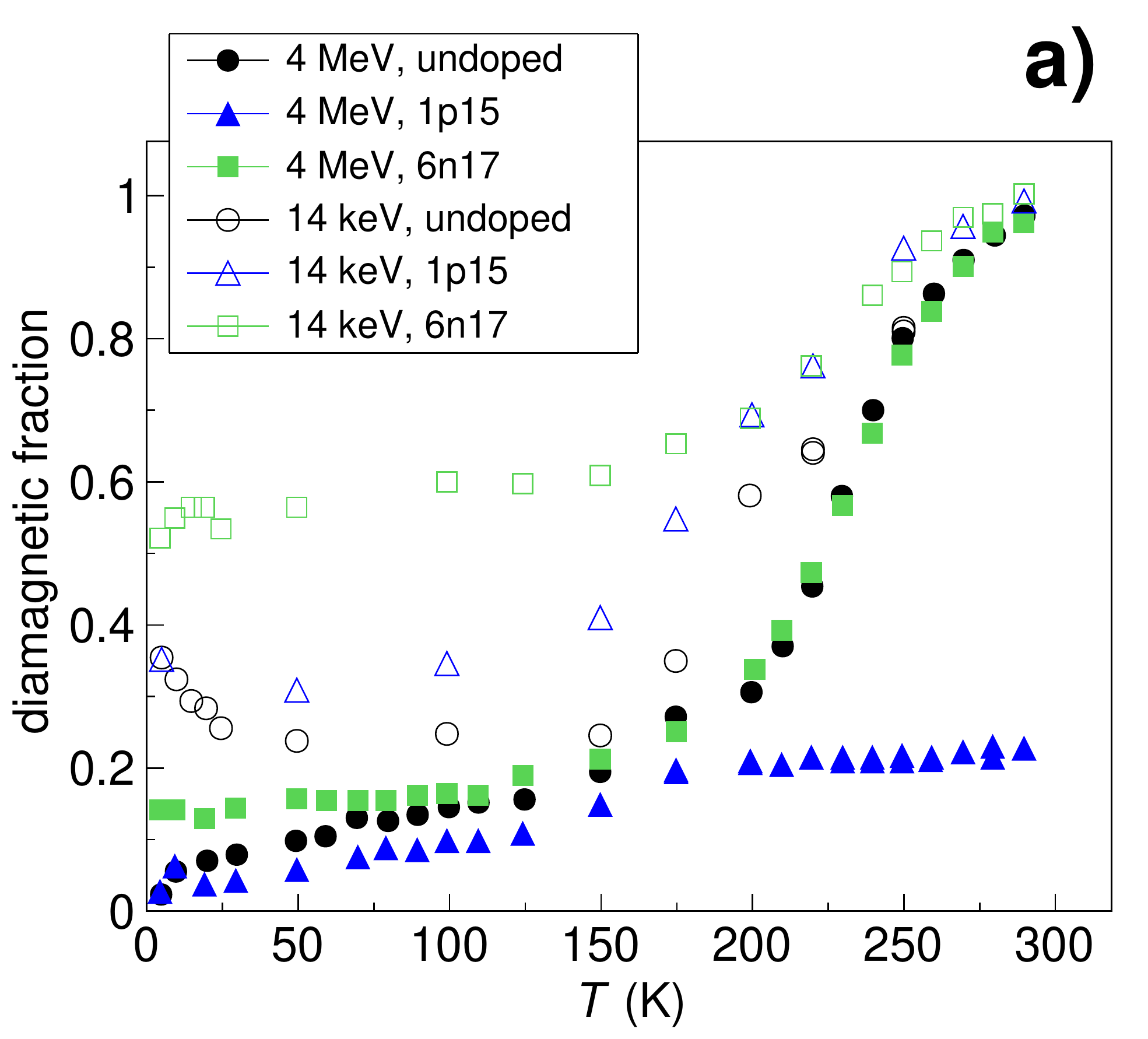}
\includegraphics[width=0.495\linewidth]{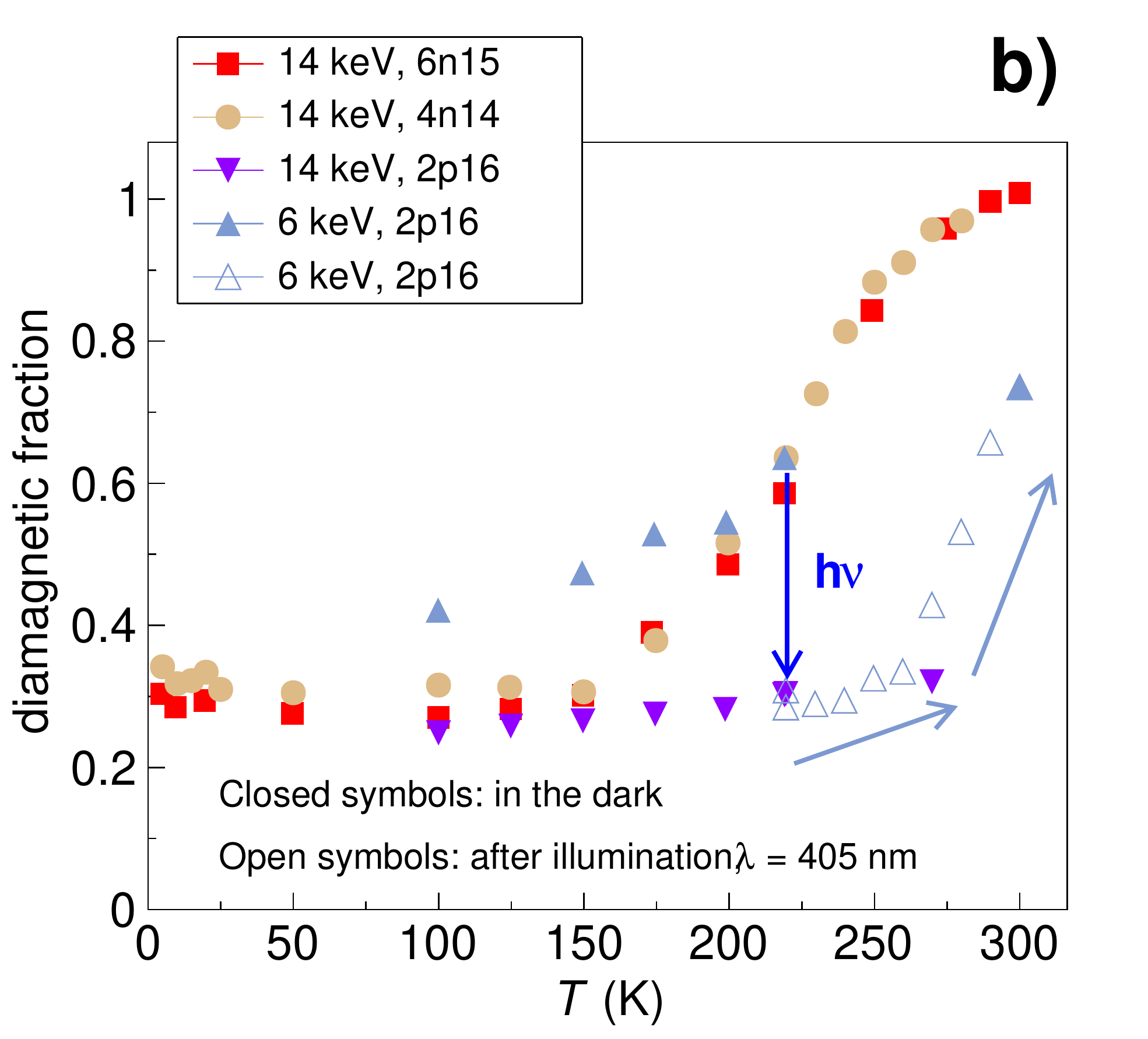}
\caption{Diamagnetic fraction as a function of temperature in Ge wafers with different doping levels: nominally undoped n-type, n-type with $n \sim 4\times 10^{14}$~cm$^{-3}$, 
$\sim 6\times 10^{15}$~cm$^{-3}$, $\sim 6\times 10^{17}$~cm$^{-3}$ (4n14, 6n15, 6n17), and
p-type with $p \sim 1\times 10^{15}$~cm$^{-3}$, $\sim 2\times 10^{16}$~cm$^{-3}$ (1p15, 2p16).
a) Comparison of LEM data (14~keV, mean depth of 80~nm, 10~mT) with DOLLY and GPS data 
(4~MeV, $\langle z\rangle \sim 300$~$\mu$m, 10~mT and 100~mT). b) LEM data (6~keV and 14~keV, 
mean depths $\langle z\rangle$~of 35 and 80~nm, respectively) for the 4n14, 6n15 and 2p16 samples. The 
arrows indicate the evolution of the diamagnetic fraction in the 2p16 sample at 6~keV after illumination with blue light at 220~K. The error bars have
about the size of the symbols and were omitted for clearer presentation of the data.}\label{Fig3}
\end{figure*}

Figure~\ref{Fig3} shows the temperature dependence of the diamagnetic fraction $F_D$ for various Ge samples with different doping levels and at different muon implantation energies,
i.e.~different mean depths of stopping $\mu^+$. $F_D$ is defined as the fraction of muons in a diamagnetic state, determined by the fraction of muons precessing at the muons' Larmor
frequency.
We begin with the description of the bulk $\mu$SR measurements of the undoped and p- and n-type samples [$p \sim 1\times 10^{15}$~cm$^{-3}$ (1p15), 
$n \sim 6\times 10^{17}$~cm$^{-3}$, (6n17)] in Fig.~\ref{Fig3}~a). The diamagnetic fraction $F_D$ is $<$~20\% at $T < 150$~K, and either slowly increasing from 5~K to 150~K for the
undoped and 1p15 sample, or nearly constant for the 6n17 sample. 
Such a weak increase of $F_D$ has been also reported in this temperature range
for the samples of Ref.~\cite{fan_influence_2008} with doping levels $\lesssim 10^{13}$~cm$^{-3}$. It can be understood as manifestation of the low-temperature formation of the \MuTm~state
through a transition state, as has been recently proposed as a general feature in the 
formation of the final muon configuration \cite{vilao_role_2017}. In the n-type sample such  an increase of diamagnetic fraction is not observed, and might be a consequence of the high electron concentration and/or a modification of the barriers in the transition state due to the doping of the lattice, favoring the formation of \MuTm~already at low temperatures. It is 
important to note that these subtle effects do not affect the dynamics at $T > 200$~K, where the dominant neutral fraction (\MuTn) becomes thermally ionized.
The faster increase of $F_D$ between 150~K and 200~K
can be attributed to the thermally activated ionization of \MuBCn, which is ``completed'' at $T \simeq 200$~K \cite{fan_influence_2008}. This appears as a flattening of $F_D$ in
the 1p15 sample: here the increase of $F_D$ due to the thermally activated formation of \MuTm~at $T > 200$~K is not observable - in contrast to the undoped and 6n17 sample - because
the presence of holes drives the reverse reaction in Eq.~\ref{eq1} too quickly. 

Implanting the muons much closer to the surface with an energy of 14~keV at a mean depth 
$\langle z\rangle \simeq$~80~nm reveals considerable differences: i), $F_D$ at $T < 150$~K is significantly larger than in the bulk, ii), the increase of $F_D$ due to thermally
activated \MuTm~formation
appears to begin at a lower temperature around 150~K, and iii), the most striking feature, we observe the thermally activated formation of \MuTm~in the 1p15 sample. The latter can
be explained by the absence of holes at least to a depth of 120~nm, i.e.~the presence of a hole depletion layer. 
This is supported by the observed weak depolarization rate at 220~K, which
is smaller than the expected depolarization rate of $\sim 0.06$~$\mu$s$^{-1}$ for $p \sim 10^{11}$~cm$^{-3}$ [Fig.~\ref{Fig1} a)], implying $p \lesssim 10^{11}$~cm$^{-3}$ in the
depletion layer. Additionally, the larger $F_D$ below 100~K indicates
an electron accumulation in the near-surface region, with $n$ ranging between $n \sim 10^{18}$~cm$^{-3}$ and 
$\lesssim 10^{19}$~cm$^{-3}$ \cite{andrianov_interaction_1978,patterson_muonium_1988}. This means that the 1p15 wafer exhibits a surface layer inversion,
where hole depletion and electron accumulation are generated by band bending at the surface.
The presence of free electrons in the accumulation region adds to the \MuTm~formation rate due to thermal activation the electron capture rate of the process  
$\rm{Mu_T^0} + e^- \rightarrow \rm{Mu_T^-}$, yielding an enhanced \MuTm~formation rate.
This explains the larger $F_D$ values at $T < 150$~K, and the lower temperature onset 
of the increase of $F_D$,
where the thermally activated \MuTm~formation sets in ($T \gtrsim 150$~K).
The larger value of $F_D$ of the 1p15 sample compared to the undoped sample
indicates a higher electron concentration or a larger band bending and accumulation of electrons in the p-type sample. The undoped sample is closer to insulating, implying weaker band bending compared to the p-type sample, i.e. a smaller electron accumulation.

The increase of $F_D$ in the undoped sample at $T < 50$~K can be explained
within the thermal spike model \cite{Vilao_thermal_2019}. Here, excess heat (due to energy liberated during the stopping process and also as a consequence of stress release to reach the final lattice configuration around the thermalized \Mu) may not be quickly enough released to the surrounding lattice due to a reduced thermal conductivity at low temperatures. This may
cause thermal ionization of the \Mu, increasing the diamagnetic fraction. In the bulk of
the sample, this effect obviously does not occur. In contrast, in the surface region with presumably larger lattice distortions \cite{Vilao_thermal_2019}, causing a reduced thermal conductivity, the thermal
spike effect can explain the increase of $F_D$.

To further support the interpretation of a hole depletion layer, we show in Fig.~\ref{Fig3} b) the results of 
a p-type sample with an order of magnitude larger hole concentration (2p16).
We also present for comparison the temperature dependencies of two n-type samples which show similar trends as the undoped sample.
In the 2p16 sample at 14 keV ($\langle z\rangle \simeq$~80~nm) the thermally activated formation of \MuTm~is no longer observed: $F_D$ does not increase at $T > 180$~K,
indicating that $p \gg 10^{14}$~cm$^{-3}$, or in other words the hole depletion layer is now significantly shifted towards the surface. This is confirmed on lowering $E_{imp}$ to 6~keV ($\langle z\rangle \simeq$~35~nm), where we again observe - as in the case of the 1p15 sample at 14 keV - the thermally activated
formation \MuTm, which means that the hole depletion layer is still present, but with reduced width. Besides hole depletion the larger value of $F_D$ below 150~K compared to
the n-type samples indicates electron accumulation.
\begin{figure*}[ht]
\includegraphics[width=0.5\linewidth]{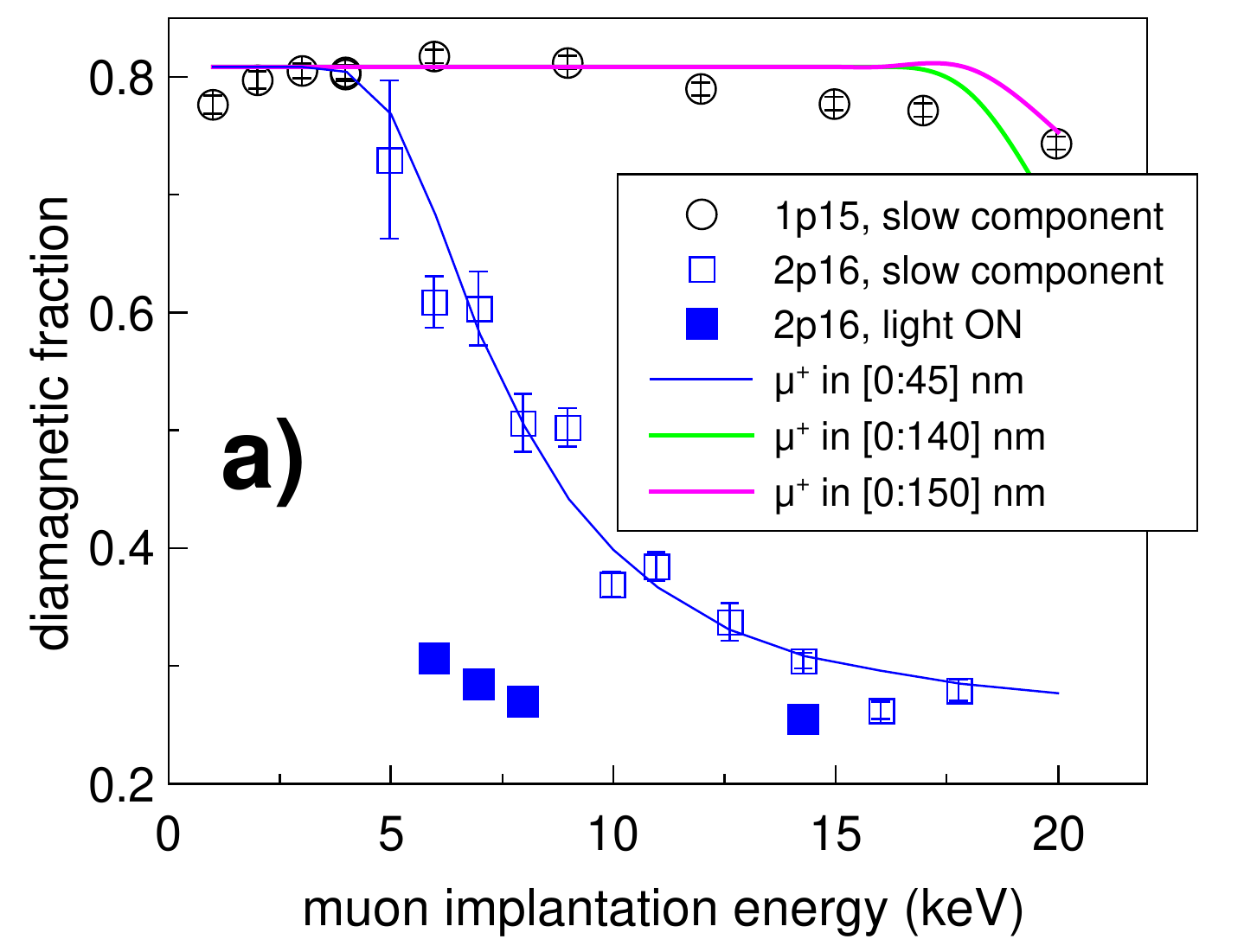}%
\includegraphics[width=0.5\linewidth]{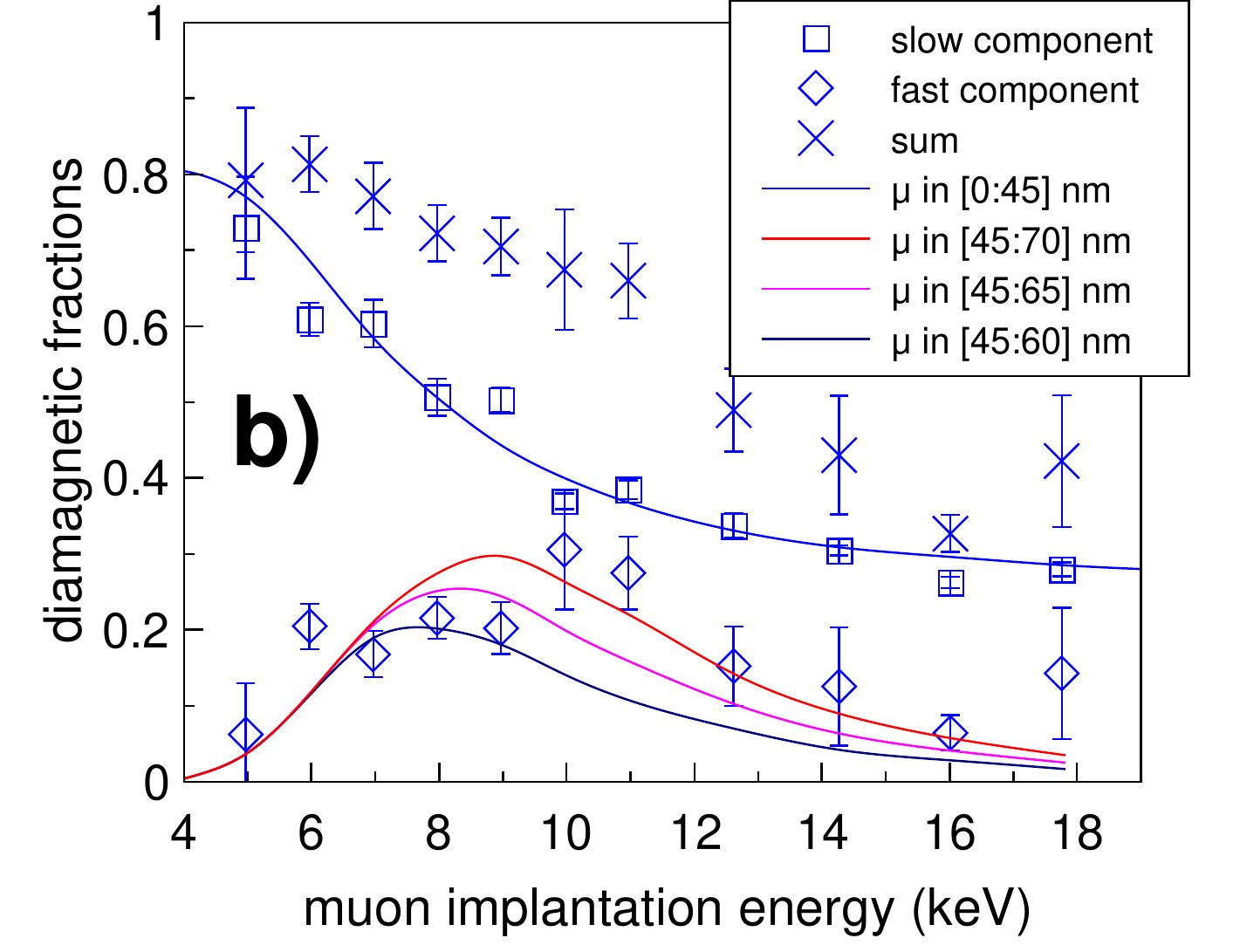}
\includegraphics[width=0.5\linewidth]{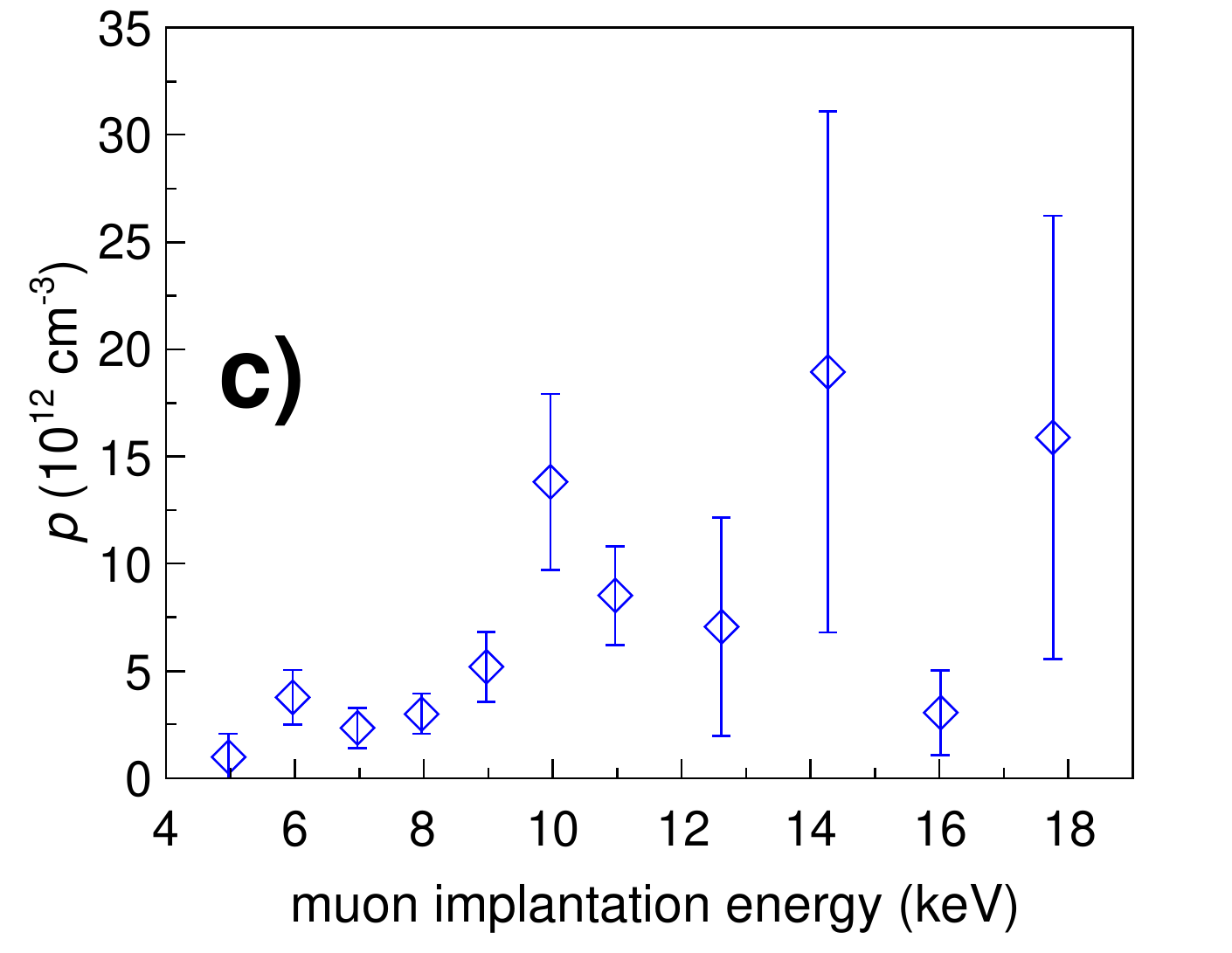}%
\includegraphics[width=0.5\linewidth]{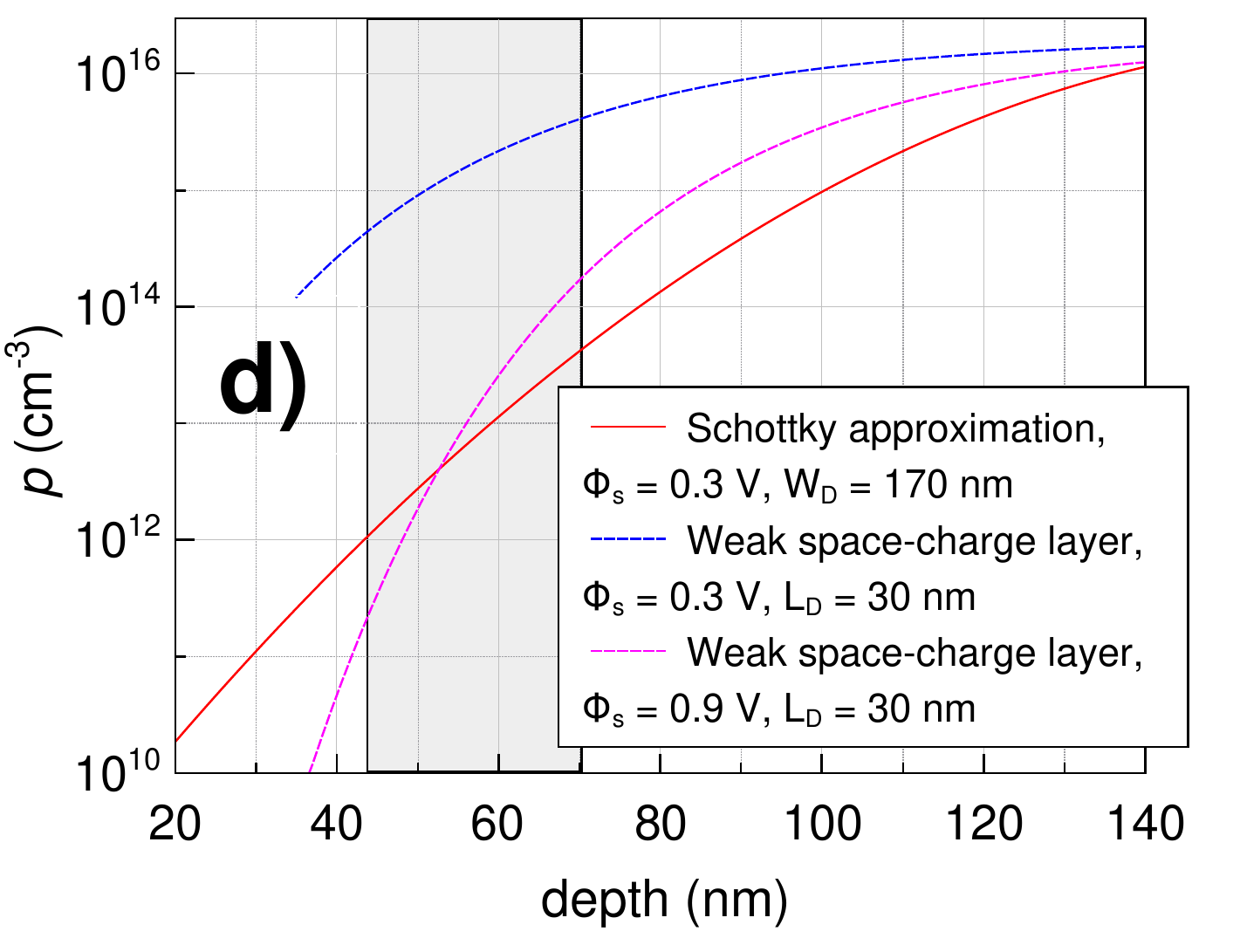}%
\caption{a) Comparison of energy dependences of the slowly relaxing diamagnetic fraction $F_{D,S}$ in p-type samples at 220~K. 
 Solid squares: $F_D$ after illumination with blue light ($\lambda = 405$~nm). The solid lines are simulations (see Appendix~\ref{appendixII}) using the calculated muon stopping profiles and 
 assuming a hole carrier concentration $p \ll 10^{12}$~cm$^{-3}$ in the regions indicated in the legend. b) 2p16 sample, where a fast component appears at
 $E_{imp} \gtrsim 5$~keV, implying the presence of holes with $p \gtrsim 3\times 10^{11}$~cm$^{-3}$. The solid lines are simulations giving the fraction of muons
 stopping in the regions displayed in the legend. c) Hole carrier concentration, calculated from $\lambda_F$ using the simulation data of Fig.~\ref{Fig1} c).
 d) Calculated $p(z)$ in the Schottky and in the weak space-charge layer approximation using Eq.~\ref{eq3}. The shaded area indicates the region where
 the fast component can be observed by LE-$\mu$SR, which means that $p$ must be in the range $\gtrsim 3\times 10^{11}$~cm$^{-3}$ and $< 10^{14}$~cm$^{-3}$ in this region.}\label{Fig4}
\end{figure*}
\subsection{Hole carrier profile in the depletion region and its manipulation by illumination}\label{SubSecManipulation}
As a first attempt to manipulate the depletion region we illuminated the 2p16 sample at 220~K with blue light ($\lambda = 405$~nm) at an intensity of up to 80~mW/cm$^{2}$. Upon light irradiation the 
value of $F_D$ measured at 220~K at 6~keV implantation energy is comparable to that measured at 14~keV implantation in the dark, indicating that the depletion region is removed, or at least significantly shifted towards the surface. After turning
off the light, $F_D$ does not change, demonstrating the persistent change/removal of the depletion layer. This effect has been observed previously\cite{prokscha_photo-induced_2013} 
and it is attributed to the trapping of photo-generated electrons in empty surface acceptor states, 
charging the surface negatively and thus pulling holes from the bulk and the photo-generated holes into the 
depletion region. On warming the sample in the dark, $F_D$ begins to increase at $T \gtrsim 270$~K, where trapped electrons from the surface acceptor
states are released and move back into the bulk of the wafer where they recombine with holes, re-establishing the hole depletion zone. The release of the electrons appears as
a thermally activated process with an energy barrier of about 1.1~eV \cite{prokscha_photo-induced_2013}. 
\begin{figure*}[ht]
\includegraphics[width=0.49\linewidth]{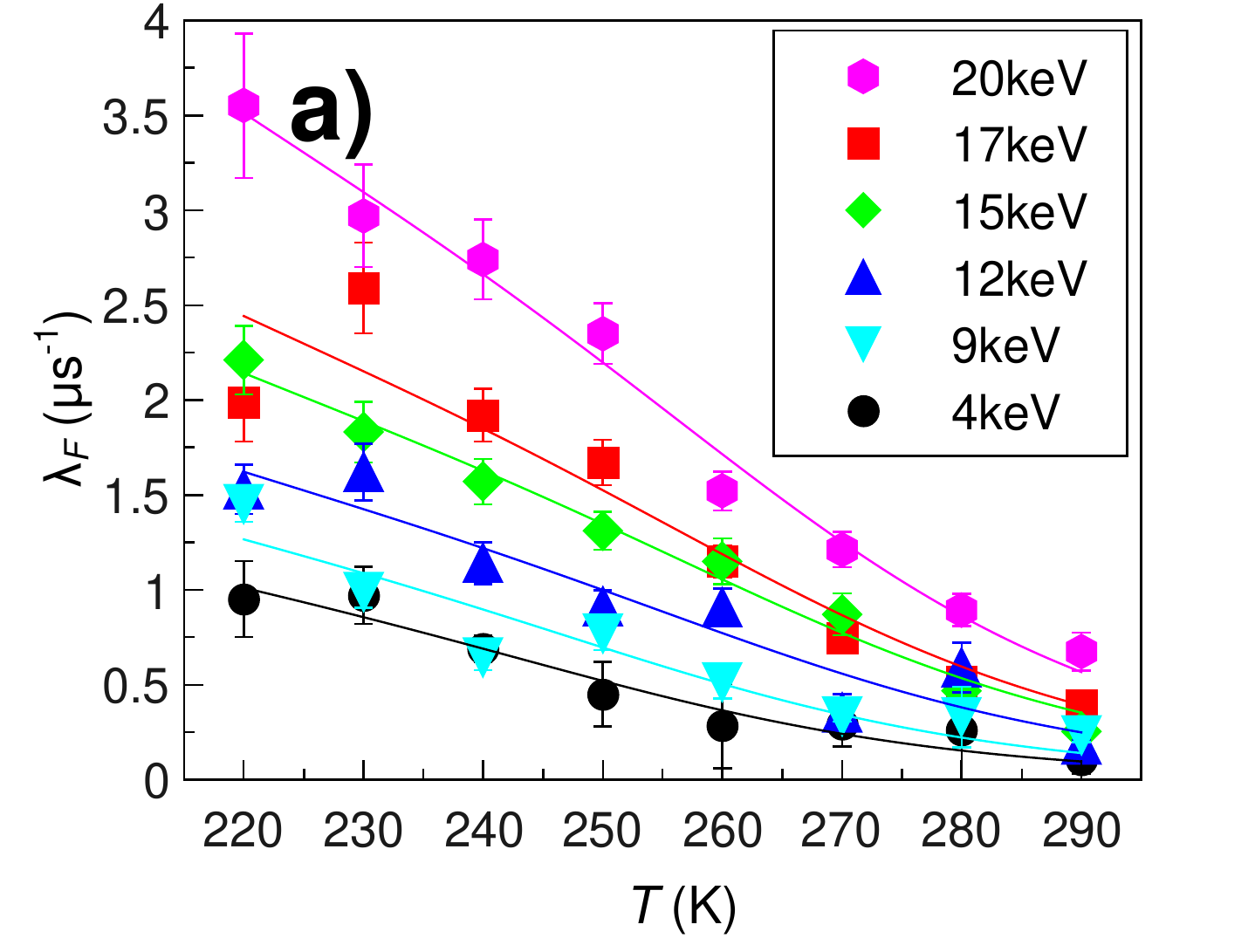}%
\includegraphics[width=0.49\linewidth]{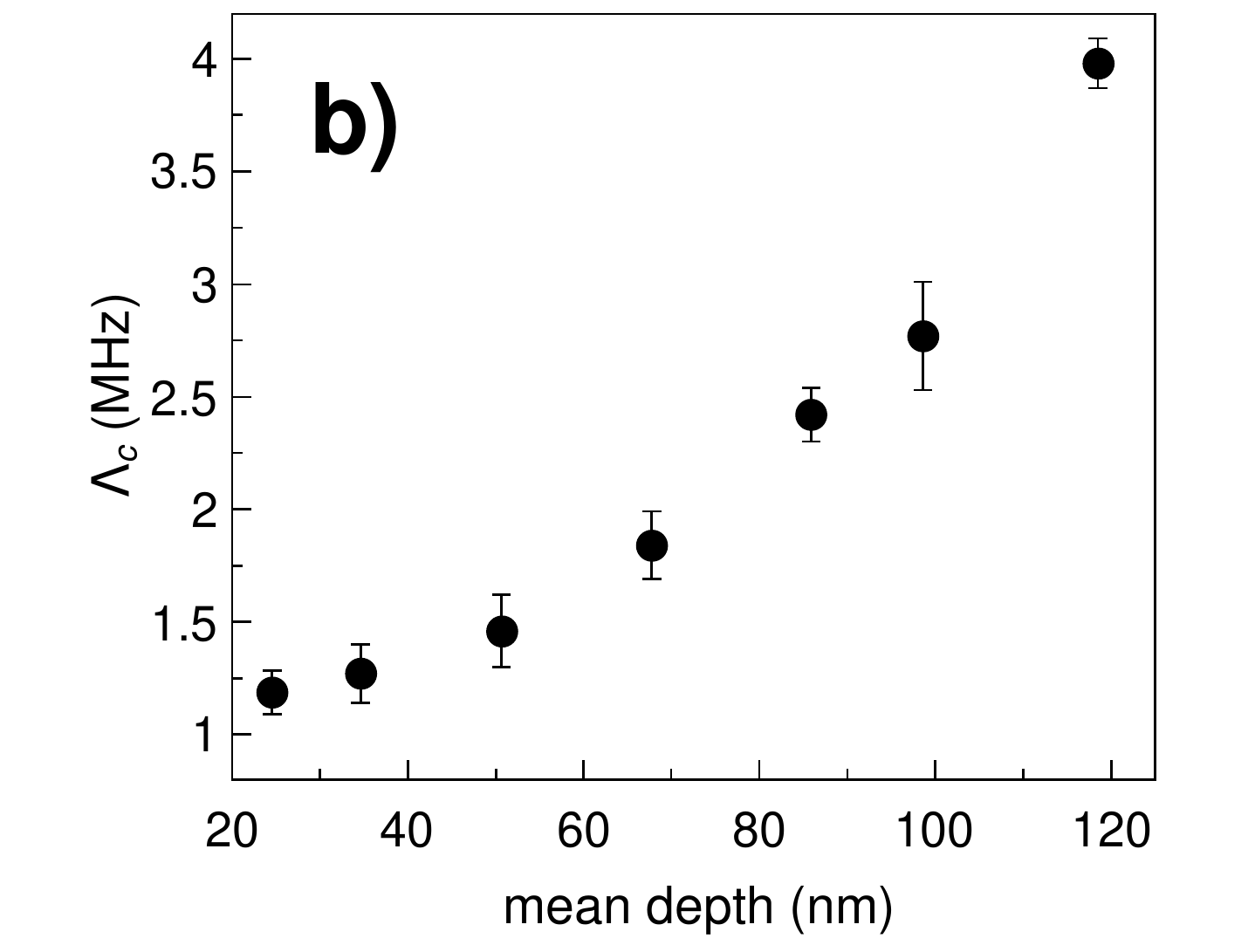}
\includegraphics[width=0.49\linewidth]{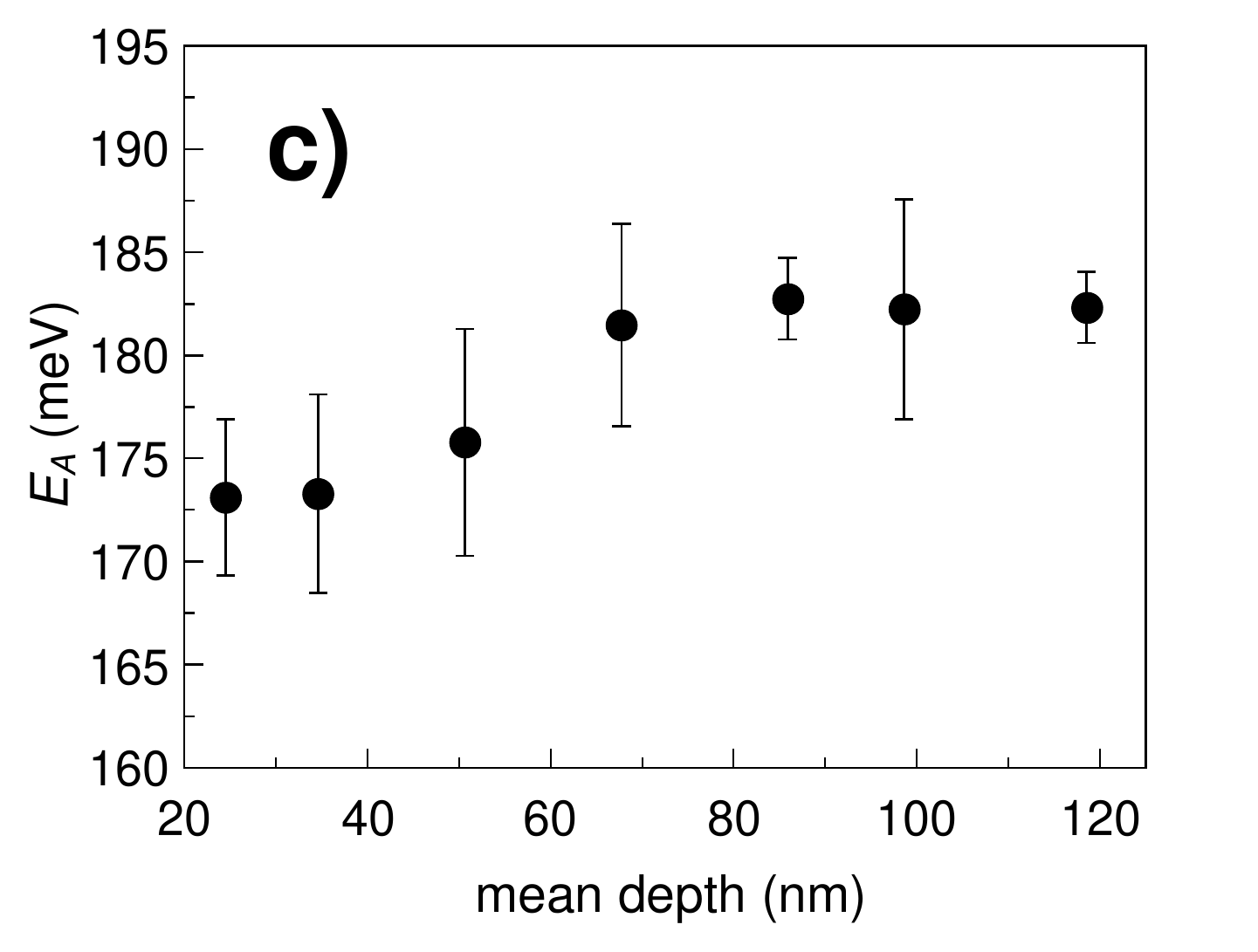}%
\includegraphics[width=0.49\linewidth]{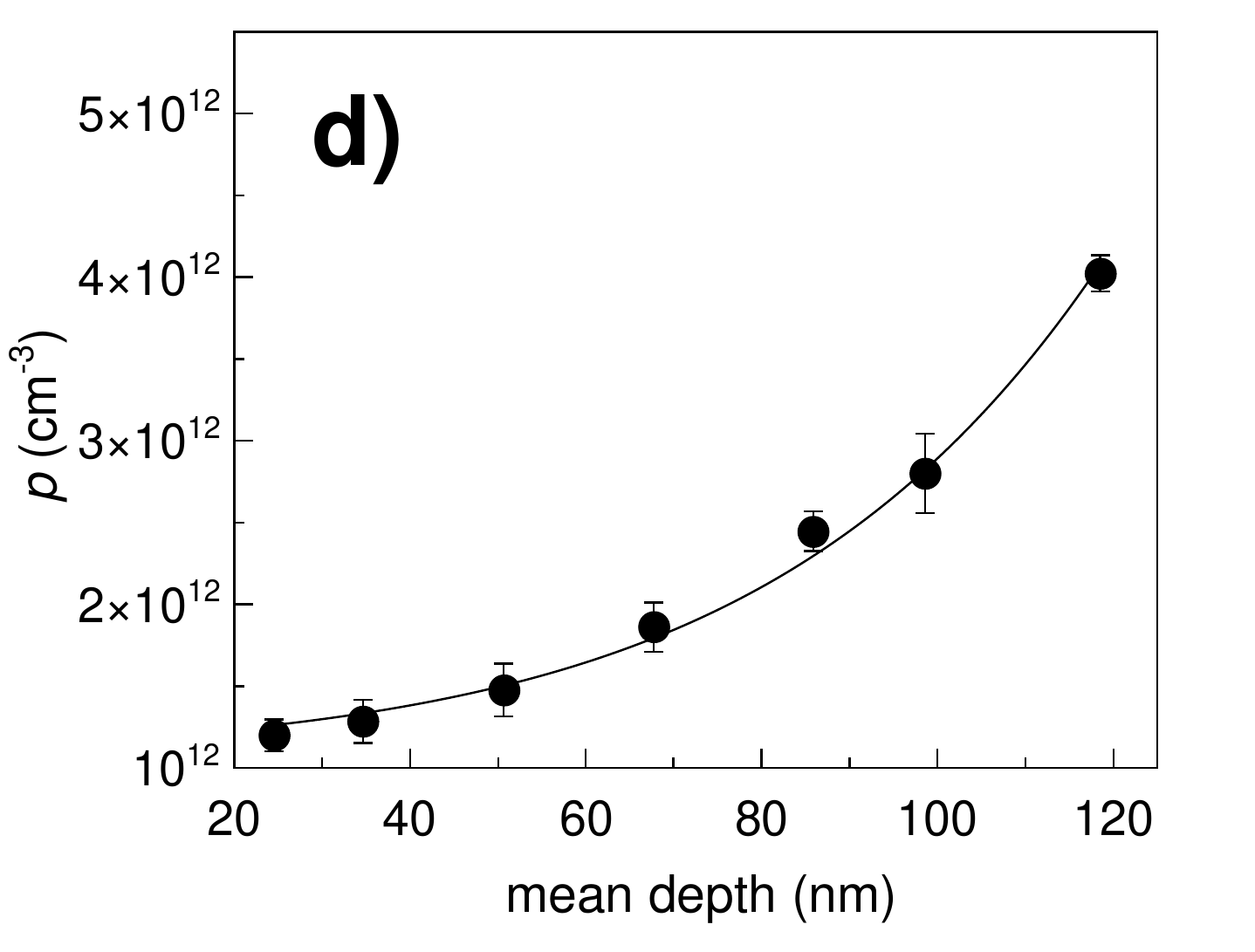}
\caption{a) Fast depolarization rate $\lambda_F$ of the 1p15 sample during illumination with red light ($\lambda = 635$~nm), 10~mT transverse field. The solid lines are 
fits of Eq.~\ref{eq2} to the data. Fit results
for the hole capture rate $\Lambda_c$ at 290 K, and the \MuTn~activation energy $E_A$ are shown in b) and c), respectively. d) Hole carrier concentration $p$ as a
function of mean depth, derived from $\Lambda_c$. The solid line is a fit of Eq.~\ref{eq3} using the Schottky approximation for $V(z)$, plus a constant offset term
$p_0$. The fit yields $W_d = 555(2)$~nm and  $p_0 = 1.1(1)\times 10^{12}$~cm$^{-3}$, where we fixed $N_A = 10^{15}$~cm$^{-3}$.}\label{Fig5}
\end{figure*}

To determine the width of the depletion region in the two p-type samples we measured $F_D$ as a function of $E_{imp}$ between 1~keV and 20~keV at $T = 220$~K, 
see Fig.~\ref{Fig4}. In the 1p15 sample only the slowly relaxing component $F_{D,S}$ is observed, indicating $p \ll 10^{12}$~cm$^{-3}$ in the entire energy/depth range accessible by low-energy muons.
The solid lines
in Fig.~\ref{Fig4}~a) are from stopping profile simulations (see Appendix~\ref{appendixII}) assuming an increase of $p$ to $\gtrsim 3\times 10^{11}$~cm$^{-3}$ beyond 140 and 150~nm, respectively. In this case,
a fast component should appear, leading to a decrease of the slowly relaxing component. No sharp drop of $F_{D,S}$ is observed, excluding an increase of $p$ in this region. 
This implies, that the depletion width $W_D$ is larger than 160~nm, the maximum range of 20-keV $\mu^+$ in Ge. The observed weak decrease of $F_{D,S}$ can be explained by a
slowly increasing activation energy $E_A$ for \MuTn~ionization as a function of depth: the presence of an electric field due to the band bending in the depletion zone will
result in a reduction of $E_A$ on approaching the surface, increasing the \MuTn~ionization rate $\Lambda_i$ at a fixed temperature, which causes an increase of 
$F_{D,S}$ \cite{prokscha_depth_2014}.
Below 5~keV both samples exhibit $F_{D,S} \simeq 0.8$. In the 2p16 sample at $E_{imp} > 5$~keV, $F_{D,S}$ begins to drop, and the data can be well described by the solid
line shown in the figure. This line is calculated assuming $p < 10^{12}$~cm$^{-3}$ at depths $z <$~45~nm, and the emergence of holes with $p >  10^{12}$~cm$^{-3}$ 
beyond 45~nm, which implies the appearance of a fast relaxing component $F_{D,F}$. Indeed, we observe this fast component, changing as a function of implantation energy 
as shown in Fig.~\ref{Fig4}~b). The data can be best modeled by assuming a depth interval of [45:70]~nm, where $p$ is in the range of  
$p \gtrsim 10^{12}$~cm$^{-3}$ and $p < 10^{14}$~cm$^{-3}$. A larger $p \gtrsim 10^{14}$~cm$^{-3}$ means a too fast depolarization of the $\mu$SR signal, 
causing a loss of the observable $F_{D,F}$. This explains the drop of the sum of both components as a function of $E_{imp}$: if the $\mu^+$ reach regions
with $p \gtrsim 10^{14}$~$cm^{-3}$, the fast component can no longer be observed leading to the reduction of $F_{D,F}$ as shown in Fig.~\ref{Fig4}~b). Figure~\ref{Fig4}~c) displays
$p$ as a function of implantation energy, derived from the measured $\lambda_F$ by scaling according to the simulation data of Fig.~\ref{Fig1} c). 
It is in the expected range, but the errors are getting large at $E_{imp} > 12$~keV due to the decreasing $F_{D,F}$ and the 
relatively poor statistics of the data. Thus, no firm conclusions about the carrier profile in the depth range [45:70]~nm can be drawn from this plot. Instead, we use 
simple modeling to calculate carrier profiles and determine the parameters of the model to obtain qualitative agreement with the experimental data. 
The hole carrier profile $p(z)$ in the depletion region depends on the local band deformation $V(z)$ at depth $z$ as \cite{lueth_solid_2010}
\begin{equation}
 p(z) = N_A \cdot \exp[-V(z)/(k_B T)], \label{eq3}
\end{equation}
where $N_A$ is the bulk acceptor density of the p-type material, and $V(z) = \phi(z) - \phi_B = \phi(z) - \phi(\infty)$, where $\phi(z)$  is the electrostatic potential with its 
bulk value $\phi_B$. Since there is actually
not only depletion but inversion at the surface, we can assume that the surface potential $\phi_s$ determining the band bending significantly exceeds $k_B T$,
$|e\phi_s| \gg k_B T$ \cite{lueth_solid_2010}. In this Schottky space-charge approximation the electrostatic potential decays quadratically from its surface value $\phi_s$ into the 
bulk following 
\begin{equation}
 V(z) = -\frac{eN_A}{2\varepsilon\varepsilon_0}(z-W_d)^2, 0 \leq z \leq W_d, \label{eq4}
\end{equation}
with a depletion width $W_d = \sqrt{2\varepsilon\varepsilon_0/(eN_A)\phi_s}$ \cite{sze_physics_2006,schroder_semiconductor_2006}. 
Choosing $W_d = 170$~nm, i.e. $\phi_s \sim 0.3$~V, and inserting in Eq.~\ref{eq3} gives the 
red curve in Fig.~\ref{Fig4}~d), resulting in a hole carrier concentration in the shaded area ($z \in [45:70]$~nm) in the range of $10^{12}$ -- $5\times 10^{13}$~cm$^{-3}$,
in good agreement with the experimental $p(z)$ in Fig.~\ref{Fig4}~c). We note, that the Schottly approximation assumes for the free carrier densities $n(z) \approx p(z) \approx 0$, 
and complete ionization of the acceptors. Due to the electron accumulation in the hole depletion zone, $n(z) > 0$, thus increasing the negative space charge in this region.
For simplicity we assumed in the calibration $n(z) < N_A$, giving a total space charge $N_A + n \simeq N_A$. This seems to be justified by the fact that the bulk $N_A$ values of
the two p-type samples seem to be the dominant densities to explain the observed differences in the depletion width estimates below.

For comparison and illustration, we show in Fig.~\ref{Fig4}~d) the blue curve $p(z)$ in the weak space-charge limit for $\phi_s = 0.3$~V 
(however, the weak space-charge limit is applicable for $|e\phi_s| < k_B T$, which is not fulfilled here)
where 
\begin{equation}
  V(z) = \phi_s \exp(-z/L_D), \label{eq5}
\end{equation}
with the Debye length $L_D = \sqrt{\varepsilon\varepsilon_0 k_B T/(e^2 N_A)} \sim 30$~nm for the 2p16 sample at 220~K. In this case, $p(z)$ would be in the range 
$5\times 10^{14}$ -- $5\times 10^{15}$~cm$^{-3}$, in contradiction to the experimental results. In order to bring $p(z)$ into the experimentally observed range
one would need to use an even larger $\phi_s = 0.9$~V (magenta curve in Fig.~\ref{Fig4}~d)), clearly outside the weak space-charge approximation.

The findings of Fig.~\ref{Fig4} can be summarized as follows. The data are well described within the Schottky approximation, implying an abrupt change of 
$p$ from $< 10^{12}$~cm$^{-3}$ to $p > 10^{13}$~cm$^{-3}$ within about 20~nm commencing at a depth of 45~nm. After illumination with blue light at 220 K, 
a persistent hole accumulation with $p > 10^{14}$~cm$^{-3}$ is established in the depletion region at implantation energies between 6~keV and 14~keV, 
corresponding to a $z$ range of $\sim 10$~nm to $\sim 120$~nm. In the 1p15 sample the width of the depletion layer is estimated to $\sim 760$~nm, $\sqrt{20}$
times larger than for the 2p16 sample (the width of the depletion layer scales with the square root of the bulk carrier concentration \cite{sze_physics_2006}). We estimate the 
space charge density $N_A\cdot W_d$ for the two samples to $\sim 3.4\times 10^{11}$~cm$^{-2}$ for the 2p16 sample, and $\sim 7.6\times 10^{10}$~cm$^{-2}$ for the 1p15 sample.

Now we turn to the manipulation of $p$ in the wide depletion region of the 1p15 sample by illumination with red light ($\lambda = 635$~nm). Under illumination, 
again a fast component appears, indicating the presence of photo-generated holes in the depletion zone.
However, after turning off the light the fast component disappears and the original slow component fraction $F_{D,S}$ is restored, meaning that 
the photo-generated holes in the depletion zone immediately disappear by recombination. This suggests, that the photo-electrons generated by red light
do not have enough energy to overcome the $\sim 1.1$~eV barrier at the surface to reach the empty surface acceptor states. In this case, 
a dynamic equilibrium of photo-generated holes and electrons is established in the depletion region, and a quick recombination with photo-generated electrons takes place after turning off
the light. As shown in Fig.~\ref{Fig5} a), the fast component can be tracked in the temperature range from 220 K to 290 K, and from close to the surface at 4~keV 
($\langle z \rangle \simeq 25$~nm) to a mean depth of 120~nm at 20 keV. This is different to illumination with blue light, where after an illumination time of 3~min with 
a laser (instead of the weaker LED source) at $\lambda = 457$~nm and $\sim 100$~mW/cm$^2$, the fast depolarization rate $\lambda_F$ exceeds values of 
60~$\mu$s$^{-1}$ ($p \gtrsim 10^{14}$~cm$^{-3}$), which is the maximum detectable depolarization rate within the experimental resolution of the LEM apparatus. 
The fast component can no longer be resolved, and it appears as a ``missing'' fraction in the $\mu$SR spectra, as in Fig.~\ref{Fig4} b). 
In contrast, the dynamic equilibrium photo-generated 
hole carrier concentration with red light is significantly smaller, with $p$ in the order of $10^{12}$~cm$^{-3}$, as we will show below. 

The increase of $\lambda_F$ as a function of energy in Fig.~\ref{Fig5}~a) points towards an increasing hole carrier concentration with increasing depth.
For a quantitative analysis we used Eq.~\ref{eq2} to fit the data (solid lines in Fig.~\ref{Fig5}~a)). In the fits we fixed according to Appendix~\ref{appendixI} i), the exponent
of the temperature dependence of $\Lambda_c$ to -2.2 [$\Lambda_c(T) = \Lambda_c(290~{\rm K})\cdot (T/290)^{-2.2}$], ii), the pre-factor 
$\Lambda_0$ in the ionization rate to $3.2\times 10^{13}$~s$^{-1}$, 
and iii), $\omega_0 = 2\pi\cdot 2150$~MHz, which is the average value of the hyperfine coupling of \MuTn~in the temperature range between 220~K and 290~K \cite{lichti_charge-state_1999}. 
The value of $\Lambda_0$ is within the range of $1.2\times 10^{13}$~s$^{-1}$ and $6.7\times 10^{13}$~s$^{-1}$ of pre-factors found in Ref.~\cite{lichti_charge-state_1999}.
The free fit parameters were $\Lambda_c(290~{\rm K})$ and the activation energy $E_A$, which are displayed in Figs.~\ref{Fig5}~b)
and c), respectively. From $\Lambda_c$ the hole carrier concentration $p(\langle z\rangle$) as a function of mean depth $\langle z\rangle$~is calculated in
Fig.~\ref{Fig5}~d), using the re-analyzed calibration data of Ref.~\cite{prokscha_simulation_2014} in Appendix~\ref{appendixI}. In doing so we assume that
the measured $\lambda_F(E_{imp})$ at implantation energy $E_{imp}$ equals $\lambda_F(\langle z\rangle)$ at the corresponding mean depth.
The justification for the validity of this assumption is discussed in Appendix~\ref{appendixIII}.

Figure~\ref{Fig5}~c) indicates a change of $E_A$ by about 10~meV under illumination between the near-surface region and a depth of about 80~nm, where it reaches its bulk value. 
This indicates a rather weak band bending under illumination with an electric field  $< 0.5$~mV/nm, where $E_A$ is reduced by $< 10$~meV \cite{prokscha_depth_2014}. 
According to Fig.~\ref{Fig5}~d), the electric field is becoming too weak to change $E_A$ within errors in regions where $p \gtrsim 2\times 10^{12}$~cm$^{-3}$.
These electric field values are confirmed by fitting Eq.~\ref{eq3} in the Schottky approximation to the data in Fig.~\ref{Fig5}~d). In order to obtain a good fit we added a
constant offset term $p_0$ to Eq.~\ref{Fig3}, where we attribute the appearance of $p_0$ to the non-equilibrium situation under illumination. The fit yields a depletion
width $W_d \sim 555$~nm, which is significantly smaller than the estimated $W_d \sim 760$~nm for the dark sample. A lowering of $W_d$ under illumination is expected due
to the reduction of band bending by the partial compensation of the surface charge by the photo-generated charge carriers. Using the reduced value of $W_d$ in Eq.~\ref{eq4} gives
an electric field of $\sim 0.5$~mV/nm at a depth of 80~nm, in good agreement with the expected value estimated above.
\section{Discussion}
%
%
%
By means of low-energy $\mu$SR we determined directly the free charge carrier profile $p(z)$ at
the surface of a semiconductor.
Knowing $p(z)$, the electrostatic potential $\phi(z)$ can be calculated using Poisson's equation.
A direct experimental determination of the bending of $\phi(z)$ at the surface is possible by 
photoemission spectroscopy. However, this method is limited to the first few nanometers at the surface, i.e. to high bulk doping levels $\gtrsim 10^{19}$~cm$^{-3}$, 
because only in this case the depletion range is comparable to the photo-electron escape depth \cite{yu_fundamentals_2010}.
We note, that $\phi(z)$ can be also determined directly by LE-$\mu$SR in cases, where a change of \Mu~activation energies is observable \cite{prokscha_depth_2014}. 
In contrast to photoemission spectroscopy the sensitivity of LE-$\mu$SR to free hole carrier concentrations is orders of magnitude larger in Ge: 
$p(z)$ as low as $\sim 10^{11}$~cm$^{-3}$ can be detected by LE-$\mu$SR over a hundred times longer length scale of about 200~nm with a resolution of a few nanometers.
This resolution exceeds significantly the capabilities of capacitance-voltage techniques, where due to the Debye-length limitation
it is not possible to profile closer than about 1$L_D$ to the surface \cite{schroder_semiconductor_2006},
which is $\sim 30$~nm in the 2p16 sample, and $\sim 130$~nm in the 1p15 sample. 

The high sensitivity to holes in the \MuTm$+h\rightarrow$~\MuTn~reaction is due to the large hole capture cross section $\sigma_c^h \sim 10^{-13}$~cm$^2$ 
of \MuTm, where we estimate $\sigma_c^h$ using the relation $\Lambda_c = p\cdot v_h \cdot \sigma_c^h$ with $\Lambda_c = 0.15$~MHz (corresponding to $p \sim 10^{11}$~cm$^{-3}$) 
as a lower detection threshold [Fig.~\ref{Fig1}~a)], and a hole velocity $v_h \sim 10^7$~cm/s.

We now turn to the implications of our results on the surface charge. The observed hole depletion and electron accumulation signifies a positively charged surface, 
suggesting the presence of empty, positively charged donor states. The change of the surface charge to negative under illumination with blue light 
requires the existence of empty surface acceptor states, which are persistently filled by photo-generated electrons. 
With red light, no persistent charging occurs, because the photo-generated electrons do not have enough energy to
overcome the surface barrier of about 1.1~eV \cite{prokscha_photo-induced_2013}. The observed electron accumulation even for the 6n17 sample means that surface 
donor states are still not filled with electrons, implying that these donor states must
be located close to the conduction band - otherwise, they would be filled and neutral. The surface acceptor states, filled under blue illumination, must also be  similarly high in energy 
as the surface donor states - otherwise, they would be filled as well at 6n17 doping, leading to a negative surface charge, and thus changing the band bending to remove 
electron accumulation. For a cleaved Ge surface without oxide layer it is well established that Fermi level pinning exists close to the valence band, causing an upward 
band bending at the Ge surface with hole accumulation and electron depletion \cite{kuzmin_origin_2016,dimoulas_germanium_2009,kuzum_characteristics_2009,gobeli_photoelectric_1964}.
This is different to our commercial Ge wafers with a $\sim$~nanometer-thin native oxide layer. The oxide can exist in various oxidation states GeO$_{\rm x}$ which may strongly
affect the electronic properities/band bending of the Ge/GeO$_{\rm x}$ interface \cite{kuzum_ge-interface_2008, schmeisser_surface_1986}. The prevailing oxidation state for native
oxide is +4 (GeO$_2$), with the presence of GeO$_{\rm x}$ with ${\rm x} < 4$ at the Ge/GeO$_2$ interface \cite{sahari_native_2011}. Assuming that the band structure at the interface 
is determined by the band alignment of GeO$_2$ with a band gap of $\sim 5.7$~eV and a conduction band offset $\Delta E_c \sim 1$~eV with respect to
Ge \cite{ohta_photoemission_2006}, we speculate that i), the band bending in Ge at the Ge/GeO$_2$ interface is opposite to a cleaved Ge surface, 
yielding electron accumulation and hole depletion as illustrated for isotype heterojunctions in
Ref.~\cite{sze_physics_2006}, and  
ii), the surface energy barrier is determined by $\Delta E_c$. While the details of the Ge/GeO$_2$ interface are important for device applications, its more detailed 
characterization is out of the scope of this study and remains for upcoming work. Here, our intention is to demonstrate the capability of charge carrier profiling in the near-surface region of a
semiconductor which allows getting insights also in the surface characteristics. 

%
%
%
%
\section{Conclusions}
%
%
%
We demonstrated that, by means of a contact-less, non-destructive local probe technique, the free charge carrier concentration profile
can be determined directly over a depth range from close to the surface up to 160~nm.
We have shown by low-energy $\mu$SR that charge carrier profiles at semiconductor interfaces can be directly studied with nanometer depth resolution, if
a muonium state forms in the semiconductor that is interacting with free carriers. The sensitivity of the technique depends on the cross section of carrier capture by the
muonium state. In the case of Ge it is the interaction of the \MuTm~state with holes at $T > 200$~K which is utilized for this purpose, where hole carrier concentrations
can be determined in the range $10^{11} - 10^{15}$~cm$^{-3}$ by the measureable effect on the muon spin depolarization rate in transverse magnetic field. This allowed us
to determine the hole carrier profile and its light induced manipulation in the hole depletion/electron accumulation region at the surface of commercial Ge wafers with a thin
native GeO$_2$ layer on top.


As an outlook, the method can be applied to characterize on a microscopic level the properties of the GeO$_{\rm x}$/Ge interface which might yield new insights for technologically
relevant Ge device applications. The study of pre-cleaned surfaces, where the native oxide layer has been removed, would be interesting to provide complementary quantitative
information on a nanometer scale of the expected hole accumulation due the Fermi level pinning close to the valence band.  

\begin{acknowledgments}
 We gratefully acknowledge the help of P. Biswas and H. Luetkens in setting up the DOLLY and GPS experiments,
 and we thank H.P.~Weber for his technical support in the operation of the LEM facility. 
 The $\mu$SR measurements were performed at the Swiss Muon Source S$\mu$S, Paul Scherrer Institut,
 Villigen, Switzerland. The $\mu$SR data were analyzed using the program {\tt musrfit} \cite{suter_musrfit_2012}.
\end{acknowledgments}

\appendix
\section{Ionization and hole capture rate parameters of the 1p15 sample}\label{appendixI}
In Ref.~\cite{prokscha_simulation_2014} we used the p-type sample with  $p \sim 10^{15}$~cm$^{-3}$ (1p15)
to determine $E_A$, $\Lambda_c(200~{\rm K})$, and the exponent $\kappa$ of the temperature dependence
of $\Lambda_c$. Longitudinal field scan data at temperatures between 220~K and 290~K were fit with 
Eq.~\ref{eq2} with fixed $\Lambda_0 = 6.7\times 10^{13}$~s$^{-1}$ and fixed hyperfine coupling constant
$A_{hfc}$ of \MuTn~at its low temperature value of 2.36~GHz.
Since $A_{hfc}$ depends on temperature we
re-analysed the data of Ref.~\cite{prokscha_simulation_2014} with a reduced $A_{hfc} = 2.15$~GHz 
in the temperature range 220~K -- 290~K,
and with a free $\Lambda_0$ parameter in the ionization rate $\Lambda_i (T)$. This procedure yields
slightly different values of the ionization and hole capture rate parameters compared to
Ref.~\cite{prokscha_simulation_2014}. These are $\Lambda_0 = 3.2(1.5)\times 10^{13}$~s$^{-1}$, $E_A = 154(12)$~meV,
$\Lambda_c(200~{\rm K}) = 2240(120)\times 10^6$~s$^{-1}$, and exponent $\kappa = -2.2(2)$ of 
the temperature dependence of $\Lambda_c = \Lambda_c(200~{\rm K}) \cdot (T/200)^\kappa$.

\section{Simulation of diamagnetic fractions}\label{appendixII}
We use a simple model to calculate the diamagnetic fractions 
in Fig.~\ref{Fig4}. The fraction $x(E, [a:b]$ of muons with implantation energy $E$, stopping in a given depth interval $[a:b]$, is calculated using the simulated muon stopping profiles (see Fig.~\ref{Fig2}). For the slow component, $F_{D,S}$ is then
given by
\begin{equation}
 F_{D,S}(E,x[0:b]) = x(E,[0:b])\cdot(0.8-0.3) + 0.3, 
\end{equation}
where we use the fact, that $F_D \sim 0.8$ at 220~K in the absence of holes, and
$F_D \sim 0.3$ is the offset diamagnetic fraction, 
which does not originate from the thermally activated ionization of \MuTn. Correspondingly, for the fast component $F_{D,F}$ in Fig.~\ref{Fig4} b) we use
\begin{equation}
 F_{D,F}(E,x[a:b]) = x(E,[a:b])\cdot(0.8-0.3), 
\end{equation}
where $(0.8-0.3) = 0.5$ is the diamagnetic fraction of muons at 220~K originating from thermally activated ionization of \MuTn, and which interact
with holes.
\section{Depolarization function of a sum of depth dependent exponential depolarization functions}\label{appendixIII}
The measured $\lambda_F = c\cdot p$, where $c$ is a constant. In the case of a depth dependent $p(z)$ the depolarization rate will become a function
of depth as well, $\lambda_F(z) = c\cdot p(z)$. This means that the depolarization function of an ensemble of muons with a stopping distribution $s(z)$ is an integral of exponentials
weighted by $s(z)$. A sum of exponentials can be approximated by a stretched exponential function \cite{johnston_stretched_2006}. However, in our case the experimental
data can be well fitted by a single exponential depolarization function. To test the validity of using a single exponential depolarization function, we added the $\mu$SR data
of energies \{4,6,9,12,15,17,20\}~keV, where the average energy is 11.9~keV. The mean depth $\langle z\rangle$ of the sum of stopping distributions is 69.3~nm, which agrees
well with the mean depth of the single 12~keV data with $\langle z\rangle = 67.5$~nm. The sum spectrum is very well fitted with a exponential depolarization function
with a reduced $\chi^2$ of 0.96 with 2036 degrees of freedom, strongly supporting the single exponential function model, 
where $\lambda_F$ of the summed data agrees within statistical error with $\lambda_F$ of the 12~keV
single energy data. The same result is obtained when choosing a different set of data, e.g. \{12,15,17,20\}~keV, where $\lambda_F$ of the sum spectra with an average energy of
16~keV equals the $\lambda_F$ of a single energy of 16~keV. This has the important implication that the measured $\lambda_F$, which is the average of $\lambda_F$'s across
the muon stopping profile, reflects  $\lambda_F$ at the mean stopping depth, i.e. $\lambda_F = \lambda_F(\langle z\rangle)$. In other words, 
$\lambda_F(\langle z\rangle) = c\cdot p(\langle z\rangle)$, justifying the interpretation of the data, that the $\lambda_F$ for a given muon implantation energy 
reflects the hole carrier concentration at the corresponding mean depth $\langle z\rangle$. 
\newpage
\bibliographystyle{apsrev4-1}

%


\end{document}